\documentclass[12pt]{amsart}
\usepackage{amssymb,amsfonts,latexsym,amscd}
\def\a{\alpha}
\def\al{a_\lambda}

\def\cnj{{n\choose j}}

\def\bra{\langle}
\def\Bra{\Big\langle}

\def\C{{\mathbb C}}

\def\E{{\mathcal E}}
\def\e{\epsilon}
\def\d{{\rm d}}

\def\Fo{{\mathcal F}}

\def\H{{\mathcal H}}

\def\ket{\rangle}
\def\Ket{\Big\rangle}

\def\N{{\mathbb{N}}}

\def\R{{\mathbb{R}}}

\def\Re{{\mathcal Re}}

\def\vac{\Omega_f}

\newcommand{\gH}{{\mathcal H}}
\newcommand{\gF}{{\mathcal F}}

\def\1{{\bf 1}}

\def\eqnn{\begin{eqnarray*}}
\def\eeqnn{\end{eqnarray*}}
\def\eqn{\begin{eqnarray}}
\def\eeqn{\end{eqnarray}}
\def\bal{\begin{align}}
\def\eal{\end{align}}

\newtheorem{theorem}{Theorem}[section]

\newtheorem{lemma}{Lemma}[section]

\newtheorem{remark}{Remark}[section]

\begin{document}

\title{{Binding conditions for atomic N-electron systems in non-relativistic QED}}

\author{Jean-Marie Barbaroux$^1$ \and
Thomas Chen$^2$ \and Semjon Vugalter$^3$}

\address{$^1$ Centre de Physique Th\'eorique, Luminy Case 907, 13288
Marseille Cedex~9, France.
jean-marie.barbaroux@cpt.univ-mrs.fr}
\address{$^2$ Courant Institute
of Mathematical Sciences, New York University, 251 Mercer Street,
New York, NY 10012-1185, USA. chenthom@cims.nyu.edu} 
\address{$^3$ Mathematik, Universit\"at M\"unchen, Theresienstrasse 39, 80333
M\"unchen, Germany.  wugalter@mathematik.uni-muenchen.de}

 \maketitle
\begin{abstract}We examine the binding conditions for atoms in non-relativistic
QED, and prove that removing one electron from an atom requires a 
positive energy. As an application, we establish the existence of a
ground state for the Helium atom.
\end{abstract}

{\em Dedicated to Professor G.   Zhislin, on
 the occasion of his seventieth birthday.}


\section{Introduction}

One of the most fundamental results in the spectral theory of
multiparticle Schr\"odinger operators is the proof of the
existence of a ground state for atoms and positive ions. It was
accomplished for the Helium atom by T.~Kato in~1951
~\cite{Kato1951}, and for an arbitrary atom by G.~Zhislin in~1960
~\cite{Zislin1960} (cf. the Zhislin theorem in
~\cite{ReedSimon1978}).

The standard approach to the proof of these results consists of
two main parts. The first key ingredient is the  HVZ - (Hunziker
-- van-Winter -- Zhislin) theorem, which establishes the location
of the essential spectrum, and gives a variational criterion for
the existence of a bound state. The latter can be referred to as
``binding conditions". The statement is that the bottom of the
essential spectrum of the whole system is defined by its
decomposition into two clusters. If the infimum of the spectrum of
the entire system is, for all nontrivial cluster decompositions,
less than the sum of the infima of the spectra of the subsystems,
it follows that the whole system possesses a ground state.

For an atom with infinite nuclear mass, this condition can be written as
\begin{equation}\label{int1}
E^V(N) < E^V(N^{\prime}) + E^0(N - N^{\prime})\quad {\rm for \
all}\quad  N^{\prime}<N,
\end{equation}
where $E^V(N)$ is the infimum of the spectrum of the atom,
$E^V(N^{\prime})$ is the infimum of the spectrum of the same atom
without $(N-N^{\prime})$ electrons, and $E^0(N-N^{\prime})$ is the
infimum of the spectrum of the system of $(N-N^{\prime})$
electrons, which do not interact with the nucleus. Obviously, in
the case of Schr\"odinger operators (in Quantum mechanics)
$E^0(N-N^{\prime})=0$, and according to the HVZ theorem, it
suffices to consider only the decompositions with $N^{\prime}=
N-1$  in~\eqref{int1}.

The second key ingredient consists of the construction of a trial state
for the Hamiltonian of the
whole atom with energy less than $E^V(N-1)$. As noted above, this
step was accomplished by T.~Kato for Helium, and by G.~Zhislin for the general case.

The problem of the existence of the ground states of atoms has
attracted new attention in the context of non-relativistic quantum
electrodynamics in the more recent literature. Bach, Fr\"ohlich
and Sigal~\cite{Bachetal1999S} first established the existence of
the ground state for the ultraviolet regularized Pauli-Fierz
Hamiltonian of an atom, for sufficiently small values of some
constants in the theory.

It was subsequently established in~\cite{Griesemeretal2001} that
the criterion for the existence of the ground state of
multiparticle Schr\"odinger operators can be extended to hold for
Pauli-Fierz Hamiltonians in non-relativistic QED, for arbitrary
values of the parameters of the theory. \footnote{A detailed
review of numerous further results connected to the existence of
ground states, mostly in Nelson-type models, can be found
in~\cite{Griesemeretal2001}. Furthermore, also cf.
\cite{Gerard2000}}

The problem, however, of devising a mathematically rigorous proof
of the fact that the binding conditions are fulfilled for atoms
apart from the one electron case, which was covered
by~\cite{Griesemeretal2001}, has turned out to be very
complicated. To clarify the main obstacles, let us recall the
basic idea underlying the proofs of the Kato and Zhislin theorems.

If the system is separated into a pair of clusters, one of which
contains $N-1$ electrons close to the nucleus, and the other
comprises a single electron far away, there is an attractive
Coulomb potential that acts on the separated particle. If the
latter is localized in a ball of radius $R$ centered at some point
with distance $bR$ from the origin, and the subsystem with $N-1$
electrons is localized in a ball of radius $R$ centered at the
origin, the intercluster Coulomb interaction can be estimated as
$CR^{-1}$ with $C<0$ for $b>N.$ At the same time, localizing the
subsystems in these balls requires an  energy $CR^{-2}$ in the
case of Schr\"odinger operators. For large $R$, the Coulomb term
is obviously dominant, and the binding condition is fulfilled.

This is contrasted by the situation in non-relativistic QED, where
the particles have to be localized together with the quantized
radiation field. One can expect, on the basis of dimensional
analysis~\cite{Griesemeretal2001}, that such a localization
requires an energy $CR^{-1}$, which makes it impossible to
establish the dominance of the Coulomb interaction by scaling
arguments.

In the work at hand, it is demonstrated how this obstacle can be
overcome. We prove that if the self-energy operator $T_0$,
restricted to states with total momentum $0$, possesses a ground
state, it is possible to construct a state consisting of an
electron coupled to a photon field, localized in a ball of radius
$R$ with energy $\Sigma_0 + o(R^{-1})$, where $\Sigma_0$ is the
self-energy of an electron. Hence, similarly as for Schr\"odinger
operators, the localization term $o(R^{-1})$ can again be
compensated by the attractive Coulomb potential. This implies that
the binding condition is fulfilled for decompositions into
clusters with $N-1$ and $1$ particles.

Existence of the ground state of $T_0$ has been recently
established for sufficiently small values of the fine structure
constant ~\cite{Chen2001}. It was proved earlier in
~\cite{Griesemeretal2001} that for the decomposition into clusters
with zero electrons and $N$ electrons, the binding condition is
also fulfilled. Thus, if an atom or a positive ion has only two
electrons, the ground state exists.

If an atom has more than two electrons, one must also verify the
binding conditions for $1 < N-N^{\prime}<N$. We note that in
contrast to the quantum mechanical case, a system of $K$ electrons
coupled to a photon field may have an energy smaller than the
self-energy of an electron multiplied by a factor $K$.

To control this case, it would be sufficient to combine a
straightforward  modification of the method developed in this
paper with a generalization of the results
of~\cite{Griesemeretal2001}, and to apply it to the case of a
system without external potential, after separating the center of
mass motion. This generalization is, however, beyond the scope of
the present work.

The first proof of the existence of the ground states for all
atoms in non-relativistic QED has, besides numerous other
important results, been accomplished by Bach, Fr\"ohlich and Sigal
in~\cite{Bachetal1999S}, by a completely different approach. To
compare the results in~\cite{Bachetal1999S} for Helium to the
results of the work at hand, we remark that the units used in our
paper correspond to those in ~\cite{Griesemeretal2001}, which
differ from the ones in~\cite{Bachetal1999S}. Furthermore, we
emphasize that while the ultraviolet cutoff in the quantized
vector potential employed in~\cite{Bachetal1999S} is, in our
units, incorporated at a value $\Lambda\sim\alpha$, where $\alpha$
denotes the fine structure constant, we are studying the
corresponding case for an ultraviolet cutoff at $\Lambda\sim 1$.
The parameter that accounts for the strength of the perturbation
produced by the photon field is in~\cite{Bachetal1999S} assumed to
be much smaller than a constant that depends on the ionization
energy of the atom, the latter being computed for the
Schr\"odinger operator of the electron subsystem. One of the key
issues in the work at hand is to devise a proof that also
encompasses the strongly nonperturbative regime, where this
parameter is allowed to be much larger than the ionization energy.
This is achieved mainly based on the parameter independence of the
results of ~\cite{Griesemeretal2001}, as well as of the methods
developed in the present paper, in addition to exploiting the
existence of the ground state of $T_0$ for small $\alpha$.


\section{Definitions and main results}\label{s2}

We consider the Pauli-Fierz Hamiltonian $H_N$ for a system of $N$
electrons in an external electrostatic potential, coupled to the
quantized electromagnetic radiation field,
\begin{equation}\label{rpf}
\begin{split}
H_N  =  \!\sum_{\ell=1}^{N} &\left\{\left(- i\nabla_{x_\ell}\otimes I_f +
\sqrt{\alpha} A_f(x_\ell)\right)^2 + \sqrt{\alpha}\sigma \cdot
B_f(x_\ell) + V(x_\ell)\otimes I_f \right\}\\
       & + \frac12\sum_{1\leq k, \ell
\leq N} W(|x_k - x_\ell|)\otimes I_f + I_{el}\otimes H_f .
\end{split}
\end{equation}
The operator $H_N$ acts on the Hilbert space
$\gH:=\gH_N^{el}\otimes \gF$, where $\gH_N^{el}$, for $N<\infty$,
is the Hilbert space of $N$ non-relativistic electrons, given by
the totally antisymmetric wave functions in $(L^2(\R^3)\otimes
\C^2)^N$, where $\R^3$ is the configuration space of a single
electron, and $\C^2$ accomodates its spin.

We will describe the quantized electromagnetic field by use of the 
Coulomb gauge condition. Accordingly, the one-photon Hilbert space is
given by $L^2(\R^3)\otimes \C^2$,
where $\R^3$ denotes either the photon momentum or configuration space, 
and $\C^2$ accounts for the two independent
transversal polarizations of the photon.  The photon Fock space is then defined
by
 $$
   \gF = \bigoplus_{n \in \N} \gF_s^{(n)} ,
 $$
where the n-photons space $\gF_s^{(n)} =
\bigotimes_s^n\left(L^2(\R^3)\otimes\C^2\right)$ is the symmetric tensor
product of $n$ copies of $L^2(\R^3)\otimes\C^2$.

We use units such that $\hbar = c = 1$, and where the mass of the
electron equals $m=1/2$. The electron charge is then given by
$e=\sqrt{\alpha}$, with $\alpha \approx 1/137$ denoting the fine
structure constant. As usual, we will consider $\alpha$ as a
parameter.

The operator that couples an electron to the quantized vector
potential is given by
\begin{equation}\nonumber
  A_f(x) = \sum_{\lambda = 1,2} \int_{\R^3}
  \frac{\zeta(|k|)}{2\pi|k|^{1/2}}
  \varepsilon_\lambda(k)\Big[ e^{ikx} \otimes a_\lambda(k)  +
  e^{-ikx} \otimes a_\lambda^\ast
  (k) \Big] \d k =: D(x) + D^\ast (x),
\end{equation}
where by the Coulomb gauge condition, ${\rm div}A_f =0$. The
operators $a_\lambda$, $a_\lambda^*$ satisfy the usual commutation
relations
 $$
  [a_\nu(k), a^\ast_\lambda(k')] = \delta (k-k') \delta_{\lambda, \nu},
  \quad [a_\nu(k), a_\lambda(k')] = 0 ,
 $$
and there exists a unique unit ray $\vac\in\gF$, the Fock vacuum,
which satisfies $a_\lambda(k)\vac=0$ for all $k\in\R^3$ and $\lambda\in\{1,2\}$.
The vectors $\varepsilon_\lambda(k)\in\R^3$ are the two
orthonormal polarization vectors perpendicular to $k$,
 $$
  \varepsilon_1(k) = \frac{(k_2, -k_1, 0)}{\sqrt{k_1^2 + k_2^2}}\qquad
 {\rm and} \qquad
   \varepsilon_2(k) = \frac{k}{|k|}\wedge \varepsilon_1(k).
 $$
The function $\zeta(|k|)$ describes the {\it ultraviolet cutoff}
on the wavenumbers $k$. We assume $\zeta$ to be of class $C^1$,
with compact support.

The operator that couples an electron to the
magnetic field $B_f = {\rm curl}A_f$ is given by
\begin{eqnarray*}
B_f (x)  &= \displaystyle\sum_{\lambda=1,2}\! \int_{\R^3}\!
\frac{\zeta_\Lambda(|k|)}{2\pi|k|^{1/2}} k\times
i\varepsilon_\lambda(k) \Big[ e^{ikx}\otimes a_\lambda(k)  +  e^{-ikx}\otimes
a_\lambda^\ast(k)\Big] \d k
\\
 & =:
K(x) + K^\ast (x).
\end{eqnarray*}
In Equation~\eqref{rpf}, $\sigma = (\sigma_1, \sigma_2, \sigma_3)$
is the 3-component vector of Pauli matrices
\begin{eqnarray*}
    \sigma_1 =
    \begin{pmatrix}
      0 & 1 \\
      1 & 0
    \end{pmatrix}\, , \ \
    \sigma_2 =
    \begin{pmatrix}
      0 & -i \\
      i & 0
    \end{pmatrix}\, , \ \
    \sigma_3 =
    \begin{pmatrix}
      1 & 0 \\
      0 & -1
    \end{pmatrix}\, .
\end{eqnarray*}
The photon field energy operator $H_f$ is given by
\begin{equation}\nonumber
H_f = \sum_{\lambda= 1,2} \int_{\R^3} |k| a_\lambda^\ast (k)
a_\lambda (k) \d k.
\end{equation}

The potentials $V$ and $W$ are relatively $-\Delta$ bounded with
relative bound zero and satisfy for positive $\gamma$, $\gamma _0$
and $r_0$ the following conditions:
\begin{equation}\label{asymptotic1}
V(x) \leq -\frac{\gamma_0}{|x|},\qquad |x|> r_0,
\end{equation}

\begin{equation}\label{asymptotic2}
W(x) \leq \frac{\gamma_1}{|x|},\qquad |x|> r_0.
\end{equation}

One of the main assumptions of the work at hand is the existence
of a ground state of the one electron self-energy operator with
total momentum $P =0$. For its precise formulation, let us
consider the case of a free electron coupled to the quantized
electromagnetic field. The self-energy operator $T$ is given by
 $$
  T = \left(- i\nabla_{x}\otimes I_f +
  \sqrt{\alpha} A_f(x)\right)^2 +
  \sqrt\alpha\sigma \cdot B_f(x) + I_{el}\otimes H_f .
 $$
We note that this system is translationally invariant, that is,
$T$ commutes with the operator of total momentum
 $$
 P_{tot} = p_{el}\otimes I_f + I_{el}\otimes P_f ,
 $$
where $p_{el}$ and $P_f = \sum_{\lambda =1,2} \int  k
a^\ast_\lambda(k) a_\lambda(k) \d k$ denote the electron and the
photon momentum operators.

Let $\H_P\cong \C^2\otimes\Fo$ denotes the fibre Hilbert space
corresponding to conserved total momentum $P$. For any fixed value
$P$ of the total momentum, the restriction of $T$ to the fibre
space $\H_P$ is given by (see e.g. \cite{Chen2001})
\begin{equation}T(P) = (P - P_f +
\sqrt{\alpha} A_f(0))^2 + \sqrt{\alpha}\sigma\cdot B_f(0) + H_f .
\end{equation}
We denote $\Sigma = \inf\sigma(T)$ and $\Sigma_0 = \inf\sigma(
T(0))$. The following assumptions will be used to formulate the
main result

\medskip
\noindent {\bf Condition $\mathfrak{C_1}$}.
\begin{itemize}
\item [i)] $\Sigma = \Sigma_0$
\item [ii)] $\Sigma_0$ is an eigenvalue of $T(0)$,
with associated eigenspace $\mathcal{E}_{\Sigma_0}$.
\item [iii)] There exists $\Omega_0 \in \mathcal{E}_{\Sigma_0}$
with a finite expectation number of photons, i.e.
 $$
   \langle N_f \Omega_0, \Omega_0 \rangle < c ,
 $$
where $N_f = \sum_{\lambda =1,2}\int a^*_\lambda(k) a_\lambda(k)
\d k$.
\item [iv)] The above eigenfunction $\Omega_0$  fulfills, for
$\lambda=1,2$ and some $p_0\in (6/5,2]$
 $$
   \| \nabla_k a_\lambda(k) \Omega_0 \| \in L^{p_0}(\R^3) + L^2(\R^3)
 $$
\end{itemize}
Condition i) was studied by Fr\"ohlich for a spinless Pauli-Fierz
model, \cite{Frohlich1974}, who proved that in this case, it is
fulfilled for all $\alpha >0$.

For the case including the $\sigma\cdot B$ term, it was proved in
\cite{Chenetal2002} that for small $\alpha$, the condition is also
fulfilled.

The existence of the eigenspace $\mathcal{E}_{\Sigma_0}$ in ii)
was recently proved for sufficiently small $\alpha$
\cite{Chen2001}, \cite{Chenetal2002}.

Finally, it will be proved in the present paper that for small
$\alpha$, the function $\Omega_0$ possesses the properties iii)
and iv). Thus, we conclude that there exists a number $\alpha_0$,
such that at least for all $\alpha \le \alpha_0$, condition
$\mathfrak{C_1}$ is fulfilled.

\medskip
The second main set of assumptions required for our analysis is
given as follows. For $M\in\N$, let $H_{M}$ denote the Pauli-Fierz
Hamiltonian for $M$ electrons defined in \eqref{rpf}.

\medskip
\noindent {\bf Condition $\mathfrak{C}_{2}$}.
\begin{itemize}
\item [i)] The operator $H_{M}$ has a ground state
 \begin{equation}\label{eq:HMgs}
  \Upsilon\in\gH =
  \gH_M^{el}\otimes\gF ,
 \end{equation}
with a finite expectation number of photons.
\item [ii)] For $\lambda=1,2$ and some $p_0\in(6/5,2]$,
 $$
  \| (I_{el}\otimes\nabla_k a_\lambda(k)) \Upsilon \|
   \in L^{p_0}(\R^3) + L^2(\R^3).
 $$
\item [iii)] Let $x_i$ $i = 1, \ldots M$ be the
position vectors of the electrons. Then,
 $$
  \left(\sum_{i=1}^M |x_i|\otimes I_f\right)\Upsilon\in \gH .
 $$
\end{itemize}


For $M\in \N$, let
 $$
  E_M = \inf\sigma(H_M) .
 $$
The main result of this article is the following
\begin{theorem}\label{mainthm}
For $N\in\N$, let the Conditions $\mathfrak{C_1}$ and
$\mathfrak{C}_{2}$ with $M=N-1$ be fulfilled, and assume that the
potentials $V$ and $W$ satisfy \eqref{asymptotic1} and
\eqref{asymptotic2}, with $\gamma_0/\gamma_1 > (N-1)$. Then,
\begin{equation}\label{mainresult1}
   E_N < E_{N-1} + \Sigma .
\end{equation}
\end{theorem}

\begin{remark}
If one assumes that the system with $M$ electrons satisfies the binding
condition of \cite{Griesemeretal2001}, it was shown in
\cite{Griesemeretal2001} that this system possesses a ground state
which satisfies all the conditions of $\mathfrak{C}_2$. In
particular, the ground state of the Hydrogen atom fulfills
$\mathfrak{C}_2$.
\end{remark}


This Theorem shows that under the above stated conditions, removing one
electron from the system costs energy. In this sense, the system is
stable with respect to the given type of ionization.

The conditions on the potential $V(x)$ and $W(x)$ cover a large
number of models in atomic and molecular physics. In particular,
for $V(x) = -\beta Z/|x|$ and $W = \beta /|x|$, the operator $H_N$
describes an atom or ion with $N$ electrons.

In the physical case, $\beta$ is equal to the Sommerfeld fine
structure constant $\alpha$. However, we would like to emphasize
that the proof of the Theorem is valid for all values of $\beta >
0$, even in the strongly nonperturbative regime $0<\beta\ll
\alpha$.

Theorem~\ref{mainthm} states that as long as the number of
electrons $N$ is less than $Z+1$ (neutral atoms and positive
ions), ionization by separation of one electron is energetically
disadvantageous.

If was earlier proved in \cite{Griesemeretal2001} that removal of
all electrons from the atom also leads to an increase of the
energy.

Combining these two results for the case $N=2$, and the binding
condition in \cite[Theorem~3.1]{Griesemeretal2001}, yields
\begin{theorem}\label{mainthm2}
The Pauli-Fierz Hamiltonian for Helium
\begin{eqnarray*}
H_2  = \!\sum_{\ell=1}^{2} &\left\{\left(- i\nabla_{x_\ell}\otimes I_f +
\sqrt{\alpha} A_f(x_\ell)\right)^2 + \sqrt{\alpha}\sigma \cdot
B_f(x_\ell) - \frac{2\alpha}{|x_{\ell}|} \otimes I_f \right\}
\\
 & + \frac{\alpha}{|x_1 - x_2|}\otimes I_f + I_{el}\otimes H_f
\end{eqnarray*}
has a ground state for all $\alpha \le \alpha _0$.
\end{theorem}

Notice that the conditions on the potential $V(x)$ require only some
type of behaviour at infinity. Therefore, instead of one nucleus with
Coulomb potential of charge $Z$, one can consider a system of nuclei
 $$
  V(x) = \sum_{i=1}^k \frac{\alpha Z_i}{|x-R_i|}
 $$
with the same total charge, in the infinite mass approximation. In
particular, for Hydrogen molecules as well as for all molecular
ions with two electrons, Theorem~\ref{mainthm} implies the
existence of a ground state for all $\alpha \le \alpha _0$.

\section{Properties of the ground state of $T(0)$.}
This section addresses the main properties of the self-energy
operator $T(0)$ that are required for the present analysis.
In particular,
existence of a ground state $\Omega_0\in\C^2\otimes\Fo$, finiteness of
the expected photon number with respect to $\Omega_0$, and
regularity of $\al(k)\Omega_0$ are discussed.

\subsection{Existence Theorem}
In the following theorem, existence
of a ground state of $T(0)$, and bounds on the associated
expected photon kinetic energy are established.

\begin{theorem}\label{Ogthm}
For $\a$ sufficiently small,
$\Sigma_0={\rm inf}\sigma(T(0))$ is a degenerate eigenvalue,
bordering to absolutely continuous spectrum, which
satisfies
$$
    |\Sigma_0|\leq c\a\;.
$$ Let $\E_{\Sigma_0}={\rm
ker}(T(0)-\Sigma_0)\subset\C^2\otimes\Fo$ denote its eigenspace.
Then, ${\rm dim}_{\C}\E_{\Sigma_0}=2$, and for any
$\Omega_0\in\E_{\Sigma_0}$, normalized by $\langle\Omega_0,\vac\rangle=1$,
the estimate $$
        \|\Omega_0\|\leq 1+ c \sqrt\a
$$
is satisfied.
Furthermore,
\eqn
        \label{AfHfOgmainest}
        \|A_f(0) \Omega_0 \| \;  , \; \|H_f^{1/2} \Omega_0 \| \leq
        c \sqrt\a
\eeqn
hold. All constants are uniform in $\a$.
\end{theorem}

For the spinless case, both results are proved in \cite{Chen2001} 
by use of the operator-theoretic renormalization group based on the smooth
Feshbach map, cf. \cite{Bachetal2002}. For the
case including spin, an outline of the proof is given in the
Appendix of \cite{Chenetal2002}, while a publication containing
the detailed proof is in 
preparation. The bound on $\|A_f(0)\Omega_0\|$ follows
straightforwardly from the one on $\|H_f\Omega_0\|$.

\subsubsection{Expected photon number}
Using Theorem~\ref{Ogthm}, we may next bound the expected photon
number with respect to $\Omega_0$.

\begin{theorem}\label{thm:chen3}
For $\a$ sufficiently small, and $\Omega_0\in\E_{\Sigma_0}$ defined as
in Theorem~\ref{Ogthm}, $\Omega_0\in{\rm Dom}(N_f^{1/2})$, where
$N_f=\sum_{\lambda=1,2}\int \al^*(k)\al(k)\d k$ is the photon
number operator, and $$
    \| N_f^{1/2} \Omega_0 \|^2 < c\sqrt\a \;.
$$
In particular,
$$
        \|\chi(|k|<1)\al(k)\Omega_0\|\leq c\sqrt\a |k|^{-1} \; .
$$
All constants are uniform in $\a$.

\end{theorem}

\begin{proof}We first remark that the integral $\int
dk\|\al(k)\Omega_0\|^2$ is ultraviolet finite, since \eqn
        \int \chi(|k|\geq 1 )\|\al(k)\Omega_0\|^2\d k &<& \int
         \chi(|k|\geq1 ) |k|
        \|\al(k)\Omega_0\|^2 \d k
        \nonumber\\
        &\leq& \langle \Omega_0 , H_f \Omega_0  \rangle
        \nonumber \\
        &\leq& c \a \;,
\eeqn using ({~\ref{AfHfOgmainest}}). We may thus assume that the
domain of the integral is the unit ball $B_1(0)$. For $|k|<1$, we
employ a similar argument as in
\cite{Frohlich1974,Bachetal1999S,Griesemeretal2001}. Using $$
        \big(:T(0): -\Sigma_0'\big)\al(k)\Omega_0
        =\big[ :T(0): \, , \, \al(k)\big]\Omega_0 \;,
$$
where $:(\,\cdot\,):$ denotes Wick ordering, and
$$
        \Sigma_0':=\Sigma_0-\langle A_f(0)^2\rangle_{\vac}
        =\inf\sigma(:T(0):)\;,
$$
we obtain
\eqn
        \al(k)\Omega_0&=&\sqrt\a R(k)\Big(  k\cdot A_f(0)
        + \frac{\zeta(|k|)}{|k|^{1/2}}\e_\lambda(k)\cdot P_f
     \nonumber\\
        &&+ \frac{\zeta(|k|)}{|k|^{1/2}}ik \wedge \e_\lambda(k) \cdot
        \sigma
        + \sqrt\a \frac{\zeta(|k|)}{|k|^{1/2}}\e_\lambda(k)\cdot
        A_f(0) \Big) \Omega_0 \;,
        \label{alkOgform}
\eeqn
where
\eqn
        R(k) := \Big( H_f + |k| + \frac{1}{2} (P_f + k)^2 - \Sigma_0'
        \Big)^{-1} \;.
\eeqn
Clearly, $\langle\vac, :T(0):\vac\rangle=0$, and
a standard variational argument shows that $\Sigma_0'<0$ for $\a>0$. Hence,
$0< R(k)   < (H_f+|k|)^{-1}$,
and
$$
        \|R(k)P_f\| \leq  \|R(k)H_f\|
        \leq 1  \; .
$$
Thus, using $\|R(k)|k|\|\leq1$ and theorem {~\ref{Ogthm}},
\eqn
        \|\chi(|k|<1)\al(k)\Omega_0\| &\leq& c \sqrt\a\chi(|k|<1) \Big(
        \|A_f(0)\Omega_0\| + 2 |k|^{-1/2}\|\Omega_0\|
            \nonumber\\
            &&\hspace{1cm}+ \sqrt\a|k|^{-1}\|A_f(0)\Omega_0\|
        \Big)\nonumber\\
            &\leq&c\sqrt\a |k|^{-1} \; .
\eeqn The right hand side is in $L^2(B_1(0))$, and the
assertion is established.
\end{proof}

For the case of a confined electron, it was proved in
\cite{Griesemeretal2001} that the
corresponding estimate exhibits
a $|k|^{-1/2}$ singularity instead of $|k|^{-1}$ as
present here, owing to
the exponential decay of the particle
wave function.

Furthermore, if the conserved momentum $P$ is non-zero, there
exists a ground state $\Omega_P(\kappa)$ for a regularized
version of the model, which includes an 
infrared cutoff below $0<\kappa\ll1$ in $A_f(0)$  (some requirements on
the cutoff function are necessary, cf. \cite{Chen2001}). 
Then, with all modifications implemented, the additional term
$$
        \sqrt\a R(k) \frac{\zeta(|k|)}{|k|^{1/2}}
            P\cdot\e_\lambda(k)
            \; \Omega_P(\kappa)
$$ enters the right hand side of ~\eqref{alkOgform}. Therefore,
$\langle\Omega_P(\kappa),N_f\Omega_P(\kappa)\rangle$ is logarithmically infrared divergent
in the limit $\kappa\rightarrow 0$,
for all $|P|>0$, and in fact, $\Omega_P(\kappa)$ does not converge to an element
in Fock space.

\subsubsection{Regularity properties of the ground state}
Next, we derive a result about the regularity of $\al(k)\Omega_0$
in momentum space, which is, in our further discussion, used for
photon localization
estimates in position space.

\begin{theorem}\label{thm:chen4}
For $\a$ sufficiently small, let $\Omega_0\in\E_{\Sigma_0}$. Then,
$$
    \|\nabla_k \al(k) \Omega_0 \|\in L^p(\R^3)+ L^2(\R^3) \;,
$$
for $1\leq p<\frac{3}{2}$.
\end{theorem}

\begin{proof}
We proceed similarly as in \cite{Griesemeretal2001}. To
begin with, we differentiate the right hand side of
\eqref{alkOgform} with respect to $k$, and observe that \eqn
        |\nabla_k R(k)| \leq (1+H_f+|k|)R^2(k)  \;,
            \label{derRkbound}
\eeqn
since $|P_f|\leq H_f$.

Let us first bound the ultraviolet
part of $\|\nabla_k \al(k)\Omega_0 \|$. For $|k|\geq1$,
\eqn
       \|\chi(|k|\geq1)\nabla_k \al(k)\Omega_0\| &=& \sqrt \a
        \|\chi(|k|\geq1)\nabla_k  R(k) k\cdot A_f(0) \Omega_0 \|
        \nonumber\\
        &\leq &  \Big( \|\chi(|k|\geq1)(1+H_f+|k|)R(k)\|+|k|^{-1}\Big)
        \nonumber\\
        &&\hspace{1cm}
        \|\sqrt \a \chi(|k|\geq1) R(k) k\cdot A_f(0) \Omega_0 \|
        \nonumber\\
        &\leq& 2\sqrt\a \|\chi(|k|\geq1)\al(k)\Omega_0\| \; ,
\eeqn and consequently, by Theorem~\ref{thm:chen3}, \eqn
        \int_{|k|\geq1}\|\nabla_k \al(k)\Omega_0\|^2 \d k\leq c \a  \; .
\eeqn
We may thus restrict our discussion to the case $|k|<1$.

Differentiating with respect to $k$, the photon polarization
vectors satisfy
\eqn
        |\nabla_k \e_{\lambda}(k)| \leq \frac{c}{\sqrt{k_1^2+k_2^2}} .
\eeqn Recalling that the cutoff function $\zeta$ is of class
$C^1$, and using Theorem~\ref{thm:chen3}, one straightforwardly
deduces that there exists a constant $c$ which is uniform in $\a$,
such that \eqn
        \| \chi(|k|<1)\nabla_k \al(k) \Omega_0 \| &\leq& c \sqrt\a
            \Big(\frac{1}{|k|^{2} }
        + \frac{1}{|k| \sqrt{k_1^2+k_2^2}}\Big) \nonumber\\
        &\leq&
        \frac{c\sqrt\a}{|k| \sqrt{k_1^2+k_2^2}} \;.
            \label{thomas2}
\eeqn
Here, one again uses $ \|R(k)P_f\| \leq  \|R(k)H_f\|\leq 1$, and
$\|R(k)|k|\|\leq1$, in addition to (~\ref{derRkbound}).
Thus, by the H\"older inequality,
\begin{equation}\label{thomas1}
\begin{split}
        \lefteqn{\Big(\int_{|k|<1}
            \| \nabla_k \al(k) \Omega_0 \|^p\d k\Big)^{1/p} } & \\
        & \ \ \ \leq \ C\sqrt\a
            \Big(\int_{|k|<1} \frac{1}{|k|^{r/2}
        (k_1^2+k_2^2)^{r/2}}\,\d k\Big)^{1/r}
        \Big(\int_{|k|<1} \frac{1}{|k|^{r^*/2}}\,\d k\Big)^{1/r^*} \;,
\end{split}
\end{equation}
with $\frac{1}{p}=\frac{1}{r}+\frac{1}{r^*}$. The integrals on the
right hand side of \eqref{thomas1} are bounded for the choices
$1\leq r^*<6$, and $1\leq r<2$, which implies that $1\leq
p<\frac{3}{2}$, corresponding to the exponent  expected from
scaling.
\end{proof}

In the case of a confined electron, \cite{Griesemeretal2001}, the bound 
analogous to (~\ref{thomas2}) is
$\frac{c\sqrt\a}{|k|^{1/2}\sqrt{k_1^2+k_2^2}}$. The reason
for the fact that it is by a factor $|k|^{1/2}$ less singular is
stated in a previous remark.
Consequently,  in  \cite{Griesemeretal2001}, the inequality corresponding to
(~\ref{thomas1}) likewise requires the choice $r<2$,
but in contrast, $r^*$ can be chosen arbitrarily large.
Therefore, the result proved in \cite{Griesemeretal2001}
holds for
$\frac{1}{p}>\frac{1}{2}+\frac{1}{\infty}=\frac{1}{2}$, that
is, $1\leq p<2$.

\section{Self-energy of localized states with total momentum $P=0$}
The goal of this chapter is to arrive at a sharp upper bound on
the infimum of the quadratic form of the operator $T(0)$, when
restricted to states where all photons are localized in a ball of
radius $R$ centered at the origin.

To this end, we recall that for the Schr\"odinger operator $-\Delta$
corresponding to a free electron, the
infimum of the spectrum on the whole space is zero, whereas the
infimum on functions supported in a ball of radius $R$, with
Dirichlet boundary conditions, is $C/R^2$.

The main result of this section is the following.
\begin{theorem}\label{thm:self-energy}
For all $R>0$, there exists a function $\Phi^R\in
\mathfrak{D}(T(0))$, such that

\noindent i) The $n$ photonic components $\Phi_n^R(y_1,\ldots,
y_n; \lambda_1,\ldots ,\lambda_n)$ fulfill
\begin{equation}\nonumber
   {\rm supp} \Phi_n^R
   \subset \{ (y_1,\ldots, y_n; \lambda_1,\ldots ,\lambda_n)\ | \
   \sup_i |y_i| <R \}
\end{equation}

\noindent ii)
\begin{equation}\label{eq:self-energy2}
 \langle T(0)\Phi^R, \Phi^R\rangle \leq
 \left( \Sigma_0 + \frac{c(R)}{R}\right)\|\Phi^R\|^2 \ ,
\end{equation}
where $c(R)$ tends to zero as $R$ tends to infinity.

\noindent iii)The function $\Phi^R$ has the following additional
properties. For all $\varepsilon>0$ and all $|x| > 2R$,
\begin{equation}\label{eq:self-energy3}
 | \langle D(x) \Phi^R, \Phi^R \rangle
 |\leq \frac{c(|x|)}{|x|} \|
 \Phi\|^2\ ,
\end{equation}
\begin{equation}\label{eq:self-energy3.5}
 | \langle D(x)^2 \Phi^R, \Phi^R \rangle
 |\leq \frac{c(|x|)}{|x|^{2}} \|
 \Phi\|^2\ ,
\end{equation}
\begin{equation}\label{eq:20.5}
 | \langle D^*(x)D(x) \Phi^R,
 \Phi^R \rangle |\leq \frac{c(|x|)}{|x|^{2}} \|
 \Phi\|^2\ ,
\end{equation}
and
\begin{equation}\label{eq:self-energy4}
 | \langle K(x) \Phi^R, \Phi^R \rangle |\leq \frac{c(|x|)}{|x|} \| \Phi\|^2
\end{equation}
where $c(|x|)$ tends to zero, uniformly in $R$, as $|x|$ tends to
infinity.
\end{theorem}
Before addressing the proof of Theorem \ref{thm:self-energy}, we
shall first demonstrate how it can be employed to construct a
state in $\mathcal{H}_1\otimes \mathcal{F}$ that accounts for a
system consisting of an electron coupled to a photonic field,
localized in a ball of radius $R$ centered at a fixed point $b$,
with energy close to the self-energy $\Sigma_0$. For that purpose,
let us, for given $x\in\R^3$, define the shift operator
$\tau_x:\mathcal{F}\rightarrow\mathcal{F}$, which, for
$\phi=(\phi_0, \phi_1, \ldots, \phi_n, \ldots )\in \mathcal{F}$,
is given by
 $$
   \tau_x\phi_n (y_1, \ldots, y_n; \lambda_1, \ldots,
   \lambda_n) = \phi_n (y_1 - x, \ldots, y_n - x; \lambda_1, \ldots,
   \lambda_n) .
 $$
\begin{theorem}\label{thm:estimate3}
Let $f$ be a real valued function in $C_0^2(\R^3)\otimes\C^2$,
supported in the unit ball centered at the origin. For
$R>0$ and $b\in\R^3$, we define
$\Theta^{R,b} \in\mathcal{H}_1\otimes \mathcal{F}$ by
\begin{equation}
 \Theta^{R,b} = \frac{f(\frac{x}{R} - b) \otimes\tau_x\Phi^R}
 {\| f(\frac{x}{R})\otimes \Phi^R \|} .
\end{equation}
Then, for all $\varepsilon>0$ and $R$ large enough independent of
$b$, we have
\begin{equation}
\left\langle \left( (i\nabla_x\otimes I_f +\sqrt{\alpha}A_f(x))^2 + \sqrt{\alpha}\sigma .
B_f(x) + I_{el}\otimes H_f \right) \Theta^{R,b}, \Theta^{R,b} \right\rangle \leq
\Sigma_0 + \frac{\varepsilon}{R} .
\end{equation}
\end{theorem}

\noindent{\it Proof of Theorem \ref{thm:estimate3}.}
For a real valued function $f$, let $f^{R,b}(x) := f(x/R -b).$ Obviously,
\begin{equation}\label{eq:25}
\begin{split}
 \Big\langle
 \left( (i\nabla_x\otimes I_f +\sqrt{\alpha}A_f(x))^2 +
  \sqrt{\alpha}\sigma . B_f(x)
 + I_{el}\otimes H_f
 \right) \Theta^{R,b}, \Theta^{R,b} \Big\rangle =\\
 \frac{1}{\|  f(\frac{x}{R} ) \otimes\Phi^R\|^2} \left(\langle
 -\Delta_x f^{R,b}, f^{R,b}\rangle \|\Phi^R\|^2 + \| f \|^2
 \langle T(0) \Phi^R, \Phi^R \rangle\right) .
\end{split}
\end{equation}
According to Theorem~\ref{thm:self-energy}, the second term on the
right hand side can be estimated by
 $$
  \frac{\| f \|^2
 \langle T(0) \Phi^R, \Phi^R \rangle}
 {\| f^{R, b}(x)\otimes \Phi^R\|^2}
 \leq \Sigma_0 + \frac{c(R)}{R}\ .
 $$
For the first term on the right hand side of \eqref{eq:25}, we have
 $$
 \frac{\langle
 -\Delta_x f^{R,b}, f^{R,b}\rangle}
 {\|  f^{R, b}(x)\|^2}
 \leq \frac{c}{R^2}\ ,
 $$
which completes the proof of the Theorem.

\subsection{Localization estimates}
In order to prove Theorem \ref{thm:self-energy}, we consider the
ground state $\Omega_0$ of the self-energy operator $T(0)$ at zero
momentum, and act on it with two spatial localization functions
$\mathcal{U^R}$ and $\mathcal{V^R}$, which constitute a partition
of unity $(\mathcal{U^R})^2 + (\mathcal{V^R})^2 = 1$ on
$\mathcal{F}$. This yields a state for which all photons are
inside the ball of radius $R$, and another state for which all
photons are outside the ball of radius $R/2$.

Clearly, the expectation of $T(0)$ with respect to $\Omega_0$ is
not equal to the sum of the expectations with respect to the two
localized states. The difference, which is usually called the
localization error, must be estimated to obtain an upper bound on
the self-energy of the localized state. In the present subsection,
we estimate the localization errors for different terms in the
operator $T(0)$.

Let us to begin with define spatial cutoff functions $u$ and $v$
as follows. We pick $u\in C_0^\infty(\R_+)$ such that
\begin{equation}\label{def-smallu}
  u(x) = \left\{
  \begin{array}{ll}
     1 & \mbox{ if } x\in [0,1/2] \\
     0 & \mbox{ if } x \geq 1
  \end{array}
  \right.\ ,
\end{equation}
$0\leq u \leq 1$ and $v:=\sqrt{1-u^2} \in C^2(\R_+)$.

For $Y= (y_1, y_2, \ldots, y_n)\in \R^n$, we denote
$\|Y\|_\infty=\max_{1\leq i\leq n} |y_i|$. For $n\in\N$ and all
$Y\in\R^n$, we also define $u_n^R(Y) = u(\frac{\| Y \|_\infty}
{R})$ and $v_n^R(Y) = \sqrt{1-u_n^R(Y)^2}$.

Next, we introduce a pair of operators $\mathcal{U^R}$ and
$\mathcal{V^R}$ on $\mathcal{F}$ by
 \begin{equation}\label{defu}
  \mathcal{U}^R\psi = \left( \psi_0, u_1^R(y_1)\psi_1(y_1), \ldots,
  u_n^R((y_1,\ldots y_n)) \psi(y_1, \ldots y_n), \ldots \right)
 \end{equation}
and
 \begin{equation}
 \mathcal{V}^R\psi = \left( \psi_0, v_1^R(y_1)\psi_1(y_1), \ldots,
  v_n^R((y_1,\ldots y_n)) \psi(y_1, \ldots y_n), \ldots \right),
 \end{equation}
where we have omitted the polarization indices from the notation.

\subsubsection{Localization error for the field energy $H_f$}
\begin{lemma}\label{lem1}
There exists $c<\infty$ such that for all $\varepsilon >0$, and all $R$
large enough,
\begin{equation}
 \langle H_f \mathcal{U}^R\psi, \mathcal{U}^R\psi \rangle + \langle
 H_f \mathcal{V}^R\psi, \mathcal{V}^R\psi \rangle - \langle H_f \psi,
 \psi \rangle \leq \langle N_f \psi, \psi \rangle \left(
 \frac{\varepsilon}{R} + \frac{c}{\varepsilon R}
 \frac{\|\mathcal{V}^{R/2} \psi\|^2}{\|\psi\|^2} \right)
\end{equation}
holds for $\psi\in \mathfrak{Q}(H_f)\cap \mathfrak{Q}(N_f)$.
\end{lemma}
\begin{proof}
Since $H_f$ maps each $n$-photon sector of the Fock space
$\mathcal{F}$ into itself, it suffices to estimate
the localization error for the $n$-photon component of $\psi$.
Furthermore, since $H_f$ acts on a function in $\mathcal{F}_s^{(n)}$
as $n |\nabla_{y_1}|$, the statement of the Lemma follows
straightforwardly from Lemma~\ref{lem2}.
\end{proof}
\begin{lemma}\label{lem2}
 There exists $c<\infty$ such that for all $\varepsilon
>0$, all $R$ large enough,
\begin{equation}
\begin{split}
 \langle  | \nabla | u(\frac{|y|}{R})\phi,
  u(\frac{|y|}{R}) \phi \rangle &+ \langle
  | \nabla |
 v(\frac{|y|}{R}) \phi, v(\frac{|y|}{R})\phi \rangle  -
 \langle | \nabla | \phi, \phi \rangle \\
 & \leq   \left( \frac{\varepsilon}{R}  +
 \frac{c}{\varepsilon R} \frac{\| \phi \chi(|y|>R)\|^2}{\|\phi\|^2} \right).
\end{split}
\end{equation}
holds for all $\phi\in C_0^\infty(\R^3)$.
\end{lemma}
\begin{proof}
By \cite[Theorem 9]{LiebYau1988}, we have
\begin{equation}
 \begin{split}
 \langle | \nabla | \phi, \phi\rangle &-
 \langle |\nabla | u(\frac{|y|}{R})\phi,
 u(\frac{|y|}{R}) \phi \rangle - \langle
 | \nabla |
 v(\frac{|y|}{R}) \phi, v(\frac{|y|}{R})\phi \rangle  \\
 &=  \frac{1}{2\pi^2}\int
 \frac{|\phi(y)|
 |\phi(z)|}{|y - z|^4}
 \left(
 \left| u(\frac{|y|}{R}) - u(\frac{|z|}{R})
 \right|^2 +
 \left| v(\frac{|y|}{R}) - v(\frac{|z|}{R})
 \right|^2 \right)
 \d y \d z .
 \end{split}
\end{equation}
Let us consider
\begin{equation}
  I  =
  \int
 \frac{|\phi(y)|
 |\phi(z)|}{|y - z|^4}
 \left| u(\frac{|y|}{R}) - u(\frac{|z|}{R})
 \right|^2 \d y \d z.
\end{equation}
The term with the function $v$ can be estimated similarly.
By symmetry, it suffices to estimate this integral in the region
where $|y| \leq |z|$. We split the integral $I$ into three parts
$I_1$, $I_2$, and $I_3$, respectively, corresponding to the regions
$\mathcal{R}_1 = \{ |z| < R/2 \}$, $\mathcal{R}_2 = \{ |z|
 > R/2 , |y - z| > R/4 \}$ and $\mathcal{R}_3 = \{ |z|  > R/2
 , |y-z| < R/4 \}$.

Since $|y| \leq |z|$, we have, in the region $\mathcal{R}_1$, $|y|
\leq |z| < R/2 $. Thus, in $\mathcal{R}_1$, we have $u(\frac{|y|}{R})
- u(\frac{|z|}{R}) = 0$. Therefore,
 $$
  I_1=0\ .
 $$
Now, for all $\varepsilon >0$
\begin{equation}
 \begin{split}
 I_2 & \leq \varepsilon \int_{\mathcal{R}_2}
 \frac{|\phi(y)|^2}{|y - z|^4} \d y \d z
 + \frac{1}{\varepsilon} \int_{\mathcal{R}_2}
 \frac{|\phi(z)|^2}{|y - z|^4} \d y \d z\\
 & \leq c \left(\varepsilon \frac{1}{R}\| \phi \|^2
 + \frac{1}{\varepsilon R}
  \| \phi\displaystyle\chi({| z | > R/2}) \|^2
\right)
 \end{split}
\end{equation}
where $c$ is a constant independent of $\varepsilon$.

Finally, since the derivative of $u$ is bounded, we have the
inequality $|u(|y|/R) - u(|z|/R)|^2 \leq c |y-z|^2 / R^2$.
This implies
\begin{equation}
\begin{split}
I_3 & \leq c \int_{\mathcal{R}_3} \frac{|\phi(y)|
|\phi(z)|}{|y-z|^4} \frac{|y-z|^2}{R^2} \d y \d z \\
 & \leq \frac{c}{R^2} \int_{\mathcal{R}_3} \frac{\varepsilon
 |\phi(y)|^2 + (1/\varepsilon)|\phi(z)|^2}{|y-z|^2} \d y \d z \\
 & \leq \frac{c \varepsilon}{R} \|\phi \|^2 +
 \frac{1}{\varepsilon R} \|\phi \chi( |z| > R/2 )\|^2 .
\end{split}
\end{equation}
\end{proof}

\subsubsection{Localization error for the operator $P_f^2$}
\begin{lemma}\label{lem3}
There exists $c<\infty$ such that for all $\varepsilon
>0$ and all $R$ large enough,
\begin{equation}
 \langle P_f^2 \mathcal{U}^R\psi, \mathcal{U}^R\psi \rangle + \langle
  P_f^2
 \mathcal{V}^R\psi, \mathcal{V}^R\psi \rangle -
 \langle P_f^2 \psi, \psi \rangle
 \leq \frac{c}{R^2} \langle
  N_f \psi, \psi \rangle
\end{equation}
holds for $\psi\in
\mathfrak{Q}(H_f)\cap \mathfrak{Q}(N_f)$.
\end{lemma}
\begin{proof}
The operator $P_f^2$ maps each $n$-photon sector into itself.
Therefore, it is sufficient to restrict the proof to
$\mathcal{F}_s^{(n)}$. We have
\begin{equation}\label{eq:lem3-1}
 \begin{split}
   \lefteqn{\langle P_f^2 u_n^R \psi_n , u_n^R \psi_n\rangle
   + \langle P_f^2 v_n^R \psi_n , v_n^R \psi_n\rangle
   - \langle P_f^2 \psi, \psi\rangle } &\\
   & = \sum_{i,j} \langle \nabla_i \nabla_j u_n^R \psi_n, u_n^R
   \psi_n\rangle +  \langle \nabla_i \nabla_j v_n^R \psi_n, v_n^R
   \psi_n\rangle -
    \langle \nabla_i \nabla_j  \psi_n,
   \psi_n\rangle \\
   & = \sum_{i,j} \langle u_n^R \nabla_i \nabla_j  \psi_n, u_n^R
   \psi_n\rangle +  \langle v_n^R \nabla_i \nabla_j  \psi_n, v_n^R
   \psi_n\rangle -
    \langle \nabla_i \nabla_j  \psi_n,
   \psi_n\rangle \\
   & + 2 \sum_{i,j} \langle (\nabla_i u_n^R) (\nabla_j \psi_n), u_n^R
   \psi_n\rangle +  \langle (\nabla_i v_n^R) (\nabla_j\psi_n), v_n^R
   \psi_n\rangle\\
   & + \sum_{i,j} \langle \psi_n \nabla_i\nabla_j u_n^R ,u_n^R
   \psi\rangle + \langle \psi_n \nabla_i\nabla_j v_n^R ,v_n^R
   \psi\rangle
  \end{split}
\end{equation}
Since $(u_n^R)^2 + (v_n^R)^2 = 1$, the first term on the right
hand side of \eqref{eq:lem3-1} is zero. Similarly, by rewriting
the second term as
 $$
   \sum_{i,j} \langle (\nabla_i (u_n^R)^2) (\nabla_j \psi_n),
   \psi_n\rangle +  \langle (\nabla_i (v_n^R)^2) (\nabla_j\psi_n),
   \psi_n\rangle
 $$
we find that it is also zero. Next, we note that $\nabla_i\nabla_j
u_n^R = 0$ and $\nabla_i\nabla_j v_n^R = 0$ if $i\neq j$, because
the functions $u$ and $v$ depend only on the $\|.\|_\infty$ norm.

Thus, we obtain
\begin{equation}
\begin{split}
  \sum_{i,j}& \langle \psi_n \nabla_i\nabla_j u_n^R ,u_n^R
   \psi\rangle + \langle \psi_n \nabla_i\nabla_j v_n^R ,v_n^R
   \psi\rangle \\
   & =
   \sum_{i} \langle \psi_n \Delta_i u_n^R ,u_n^R
   \psi\rangle + \langle \psi_n \Delta_i v_n^R ,v_n^R
   \psi\rangle \\
   & = n  \langle \psi_n \Delta_1 u_n^R ,u_n^R
   \psi\rangle + \langle \psi_n \Delta_1 v_n^R ,v_n^R
   \psi\rangle \\
   & \leq n \frac{c}{R^2} \|\psi_n\|^2\ ,
   \end{split}
\end{equation}
where in the last inequality, we used that for some constant $c$,
we have $|\Delta u_n^R| \leq c R^{-2}$ and $|\Delta v_n^R| \leq c
R^{-2}$.
\end{proof}

\subsubsection{Localization error for $P_f A_f(0)$}
\begin{lemma}\label{lem4}
Let $\psi\in \mathfrak{Q}({P_f
A_f(0)})\cap\mathfrak{D}(P_f)\cap\mathfrak{D}(N_f)$, and assume that for some
$p_0\in (6/5,2]$,
 $$
  \|\nabla_k a_\lambda(k) \psi\|_{\gF} \in L^{p_0}(\R^3) + L^2(\R^2).
 $$
Then, the inequality
\begin{equation}
 \Big|\langle P_f A_f(0) \mathcal{U}^R\psi, \mathcal{U}^R\psi \rangle + \langle
  P_f A_f(0)
 \mathcal{V}^R\psi, \mathcal{V}^R\psi \rangle - \langle P_f A_f(0) \psi, \psi
 \rangle\Big|
  \leq \frac{c}{R^{1+\delta}}
\end{equation}
holds with $\delta = (p_0-6/5)/2$.
\end{lemma}
\begin{proof}
Throughout this proof, we will write $\int \d y$ for integration
over the $y$ variable, and summation over the polarization $\lambda$.
Here and in the rest of the paper, we define $G_\lambda(x)$ as the
Fourier transform of the vector function
 $$
   \frac{\varepsilon_\lambda(k)}{|k|^\frac12} \zeta(k) .
 $$
In addition, everywhere where it does not lead to any
misunderstanding, we will omit the photon polarization index $\lambda$.

We have
\begin{equation}
\begin{split}
 \lefteqn{\langle P_f D(0) \mathcal{U}^R\psi, \mathcal{U}^R\psi
  \rangle + \langle P_f D(0) \mathcal{V}^R\psi, \mathcal{V}^R\psi
  \rangle - \langle P_f D(0) \psi, \psi \rangle} & \\ & = \!i\!
  \sum_n\!\sqrt{n+1}\Bigg\{ \!\! \int\!\! G(- y_{n+1}) \psi_{n+1}
  \!\sum_{i=1}^n \overline{(\nabla_i \psi_n)} \left( u_{n+1}^R u_n^R\!
  +\! v_{n+1}^R v_n^R\! -\! 1\right) \d y_1 \ldots \d y_{n+1}\\ & \ \
  \ \ \ \ \ \ \ \ \ + \int G(- y_{n+1}) \psi_{n+1} \overline{\psi_n}
  \left(\sum_{i=1}^n u_{n+1}^R\nabla_i u_n^R + v_{n+1}^R\nabla_i v_n^R
  \right) \d y_1\ldots \d y_{n+1} \Bigg\} \\ & =: \sum_n ( a_n + b_n
  )\ .
\end{split}
\end{equation}
We first estimate the term $a_n$.  We denote $F = u_{n+1}^R u_n^R
+ v_{n+1}^R v_n^R -1$. For $|y_{n+1}|\leq R/2$, either $\| Y
\|_\infty = |y_{n+1}|$ and then $u_{n+1}^R (Y) = u_n^R(y_1,\ldots,
y_n) =1$ and  $v_{n+1}^R (Y) = v_n^R(y_1,\ldots, y_n) =1$, or $\|
Y \|_\infty = |y_{k}|$, for some $k\neq n+1$, and then
$u_{n+1}^R(Y) = u_n^R(y_1, \ldots, y_n)$ and  $v_{n+1}^R(Y) =
v_n^R(y_1, \ldots, y_n)$. In both cases, we get $F=0$. Thus for
$\delta>0$ sufficiently small, we have
\begin{equation}
\begin{split}
 |a_n|   = & \left| \sqrt{n+1}\!\! \int_{|y_{n+1}|\geq
 R/2}\!\! G(- y_{n+1}) \psi_{n+1} \!\sum_{i=1}^n \overline{(\nabla_i
 \psi_n)} F \d y_1 \ldots \d y_{n+1}\right| \\ & \leq \sqrt{n+1}\!\!
 \int_{|y_{n+1}|\geq R/2}\!\! (1+|y_{n+1}| )^{1-\delta} |G(- y_{n+1})|
 |\psi_{n+1}|(1+|y_{n+1}|)^{2\delta}\\ & \times
 \frac{1}{(1+|y_{n+1}|)^{1+\delta}} (P_f\psi)_n \d y_1 \ldots \d
 y_{n+1}\\  \leq& \!\frac{1}{R^{1+\delta}}\! \int\!\!  \sqrt{n+1}
 |\psi_{n+1}|(1+|y_{n+1}|)^{2\delta} (1+|y_{n+1}| )^{1-\delta} \\
 &|G(-
 y_{n+1})| |(P_f\psi)_n| \d y_1 \ldots \d y_{n+1}
\end{split}
\end{equation}
Applying the Schwarz inequality, we arrive at
\begin{equation}
\begin{split}
 |a_n|   \leq & \frac{2^{1+\delta}}{R^{1+\delta}} \|\sqrt{n+1} \psi_{n+1} (1+
 |y_{n+1}|)^{2\delta}\|_{L^2_{n+1}}~\|(1+
 |y_{n+1}|)^{1-\delta} G\|_{L^2(\d y_{n+1})} \\
 &\|(P_f \psi)_n\|_{L^2_n},
\end{split}
\end{equation}
where for brevity, $L^2_k:=L^2(\d y_1,\dots,\d y_k)$.
According to Lemma~\ref{lem-appendix1} in the Appendix, one finds that $\|(1+
 |y_{n+1}|)^{1-\delta} G\|_{L^2(\d y_{n+1})}$ is finite. Therefore,
\begin{equation}\label{eq:lem3-4}
\begin{split}
\sum_n |a_n| & \leq
  \frac{c}{R^{1+\delta}} \sum_n \left(
 \|\sqrt{n+1} \psi_{n+1} (1+
 |y_{n+1}|)^{2\delta}\|^2_{L^2_{n+1}} + \|(P_f \psi)_n\|^2_{L^2_n}
 \right)\ .
 \end{split}
\end{equation}
We note that
 $$
  \| \nabla_k a_\lambda(k) \psi\|_{\gF} \in L^{p_0}(\R^3,\d k) + L^2(\R^3,\d k)
 $$
implies
 $$
  \sum_n (n+1) \|\psi_{n+1}(y,.)\|^2_{L^2_n} (1+ |y|)^2
  \in L^{q_0/2}(\R^3, \d y)+L^{1}(\R^3, \d y)\  ,
 $$
with $\frac{1}{p_0}+\frac{1}{q_0}=1$, by the Hausdorff-Young inequality.
Consequently, one can straightforwardly verify that for $\delta = (p_0 - 6/5)/2$,
\begin{equation}\label{eq:lem3-5}
 \sum_n
 \|\sqrt{n+1} \psi_{n+1} (1+
 |y_{n+1}|)^{2\delta}\|^2_{L^2_{n+1}} < c\ .
\end{equation}
Moreover,
\begin{equation}\label{eq:lem3-6}
 \sum_n \|(P_f \psi)_n\|^2_{L^2_n} <c\ ,
\end{equation}
since $\psi \in \mathfrak{D} (P_f)$. Inequalities
\eqref{eq:lem3-4}-\eqref{eq:lem3-6} imply that
\begin{equation}\label{eq:lem3-7}
 \sum_n |a_n| \leq \frac{c}{R^{1+ \delta}}\ .
\end{equation}
Let us turn to the estimate of $b_n$. If $\max_{i=1,\ldots,n}
|y_i| \neq |y_{n+1}|$, then
 $$
   \sum_{i=1}^n\!\Big( u^R_{n+1}(y_1,\ldots y_{n+1}) \nabla_i
    u^R_n(y_1,\ldots y_{n}) +  v^R_{n+1}(y_1,\ldots y_{n+1})
     \nabla_i v^R_n(y_1,\ldots y_{n})\Big)\! =\! 0\ .
 $$
If  $\max\{ y_1,\ldots, y_{n+1}\} = |y_{n+1}|$, then $\nabla_i
u^R_n = \nabla_i v^R_n =0$ for all $(y_1,\ldots y_{n})$, such that one finds
$\max_{k=1,\ldots, n} |y_k| > |y_i|$. This means that except on a
set of measure zero in $\R^n$, the functions $ u^R_{n+1}\nabla_i
u^R_n +  v^R_{n+1} \nabla_i v^R_n$ have disjoint supports.
Therefore,
 $$
  \sum_{i=1}^n  u^R_{n+1}\nabla_i
  u^R_n +  v^R_{n+1} \nabla_i v^R_n \leq \frac{c}{R}\ .
 $$
Moreover, $\nabla_i u^R_N$ and $\nabla_i v^R_N$ have support in
the set $\{ |y_i| \in [R/2, R] \}$, thus, since from the above, we
only have to consider the region where $|y_{n+1}| >
\max_{i=1,\ldots,n} |y_i|$, we get $|y_{n+1}| > R/2$, hence
\begin{equation}\label{eq:lem3-8}
\begin{split}
 |b_n| & \leq \frac{c}{R} \sqrt{n} \int_{|y_{n+1}| >R/2}
 |G(-y_{n+1})|~|\psi_{n+1}|~|\psi_n| \d y_1\ldots \d y_{n+1} \\ & \leq \frac{c}{R}
 \int_{|y_{n+1}| >R/2} (1+|y_{n+1}|)^{-1/2}
 |G(-y_{n+1})|(1+|y_{n+1}|)^{1/2} \\
 &\times|\psi_{n}| \sqrt{n}|\psi_{n+1}| \d
 y_1\ldots \d y_{n+1}\ .
\end{split}
\end{equation}
Applying the Schwarz inequality and Lemma~\ref{lem-appendix1}, we obtain
from \eqref{eq:lem3-8}
\begin{equation}\label{eq:lem3-9}
 \sum_n |b_n| \leq
 \frac{c}{R^{3/2}} \left(\|\psi\|^2 + \|N_f \psi\|^2
 \right) .
\end{equation}
Inequalities \eqref{eq:lem3-7} and \eqref{eq:lem3-9} complete the
proof of Lemma~\ref{lem4}.
\end{proof}
\subsubsection{Localization error for $A_f(0)^2$}
\begin{lemma}\label{lem5}
Let $\psi\in \mathfrak{Q}({ A_f(0)^2})\cap\mathfrak{D}(N_f)$, and let
for some $p_0\in (6/5, 2]$
 $$
  \|\nabla_k a_\lambda(k) \psi\| \in L^{p_0}(\R^3) + L^2(\R^3).
 $$
Then, the inequality
\begin{equation}
 \langle A_f(0)^2 \mathcal{U}^R\psi, \mathcal{U}^R\psi \rangle + \langle
  A_f(0)^2
 \mathcal{V}^R\psi, \mathcal{V}^R\psi \rangle - \langle A_f(0)^2 \psi, \psi
 \rangle
  \leq \frac{c}{R^{1+\delta}}
\end{equation}
holds with $\delta = (p_0 -6/5)/2$.
\end{lemma}
\begin{proof}
Using the canonical commutation relations, we have
\begin{equation}\nonumber
\begin{split}
 A_f(0)^2 = D(0)^2 + D^*(0)^2 + 2\Re D^*(0)D(0) + c I ,
\end{split}
\end{equation}
where the constant $c$ depends on the ultraviolet cutoff.
Therefore, it is sufficient to compute the localization error for
$D(0)^2$ and $D^*(0) D(0)$. We have
\begin{equation}\label{eq:lem5-1}
\begin{split}
 \langle D(0)^2 \mathcal{U}^R\psi,
 \mathcal{U}^R\psi \rangle + \langle
 D(0)^2 \mathcal{V}^R\psi, \mathcal{V}^R\psi
 \rangle - \langle D(0)^2 \psi, \psi \rangle \\
 = \sum_n \sqrt{n+1}\sqrt{n+2} \int G(y_{n+2}) G(y_{n+1})
 \psi_{n+2} \overline{\psi_n} \\
 \times\left( u^R_{n+2}u^R_n +
 v^R_{n+2} v^R_n - 1\right) \d y_1\ldots \d y_{n+2}
\end{split}
\end{equation}
In the region where $\max_{i=1,\ldots, n+2}|y_i| \neq \max\{
|y_{n+1}|, |y_{n+2}| \}$, we find
\begin{equation}\label{eq:lem5-2}
  \left( u^R_{n+2} u^R_n +  v^R_{n+2} v^R_n - 1\right)(y_1,\ldots,
  y_{n+2}) = \left( (u^R_n)^2 +  (v^R_n)^2 - 1\right)(y_1,\ldots,
  y_{n}) = 0\ .
\end{equation}
In the region where $\max_{i=1,\ldots, n+2}|y_i| = |y_{n+2}|\leq R/2$,
we have
$$
    u^R_{n+2}(y_1,\ldots, y_{n+2}) = u^R_{n}(y_1,\ldots, y_{n})=1
$$ and
$$
    v^R_{n+2}(y_1,\ldots, y_{n+2}) = v^R_{n}(y_1,\ldots, y_{n}) = 0 .
$$
This yields \eqref{eq:lem5-2} in that case. Similarly, in the
region where $\max_{i=1,\ldots, n+2}|y_i| = |y_{n+1}|\leq R/2$, equation
\eqref{eq:lem5-2} holds. Therefore, in \eqref{eq:lem5-1}, it suffices to
carry out the integration in the region $\{ (y_1, \ldots y_{n+2})\ |\ |y_{n+1}|\geq
R/2\} \cup \{ (y_1, \ldots y_{n+2})\ |\ |y_{n+2}|\geq R/2\}$. Let us
consider the integral in the first region. The other will be treated
the same way. We have
\begin{equation}\label{eq:50}
\begin{split}
 \left|\sqrt{n+2}\sqrt{n+1}\!\! \int\!\!  G(y_{n+2}) G(y_{n+1}) \psi_{n+2}
 \overline{\psi_n} \left( u^R_{n+2}u^R_n + v^R_{n+2} v^R_n - 1\right)
 \d y_1\ldots \d y_{n+2}\right| \\ \leq \frac{2^{1+\delta}}{R^{1+\delta}}
 \int |G(y_{n+1})| (1 + |y_{n+1}| )^{1-\delta} |G(y_{n+2})|\sqrt{n+1}
 |\psi_n| \\ \ \ \ \times\sqrt{n+2}|\psi_{n+2}|(1+
 |y_{n+1}|)^{2\delta} \d y_1\ldots \d y_{n+2}\ .
\end{split}
\end{equation}
Applying the Schwarz inequality and using \eqref{eq:lem3-5} as in
Lemma~\ref{lem4}, we obtain the estimate
\begin{equation}\nonumber
 \begin{split}
 \Big|\sum_n\!\!\sqrt{n+1}\sqrt{n+2}\!\! \int\!\!\! G(y_{n+2}) G(y_{n+1})
 \psi_{n+2} \overline{\psi_n}\!& \\
 \left( u^R_{n+2}u^R_n\! +\!
 v^R_{n+2} v^R_n\! -\! 1\right)\! \d y_1\ldots \d y_{n+2}\Big|
 &\!\leq\! c\frac{1}{R^{1+\delta}} .
 \end{split}
\end{equation}
We have
\begin{equation}\nonumber
\begin{split}
\langle D^*(0)D(0) \mathcal{U}^R\psi,
 \mathcal{U}^R\psi \rangle + \langle
 D^*(0)D(0) \mathcal{V}^R\psi, \mathcal{V}^R\psi
 \rangle - \langle D^*(0)D(0) \psi, \psi \rangle \\
 = \sum_n (n+1) \int G(y_{n+1}) \overline{G(z_{n+1})}
 \psi_{n+1}(y_1,\ldots, y_n, y_{n+1})
 \overline{\psi_{n+1}}(y_1,\ldots, y_n, z_{n+1})\\
 \times \bigg( u^R_{n+1}(y_1,\ldots, y_n, y_{n+1})
 u^R_{n+1}(y_1,\ldots, y_n, z_{n+1}) \\
 + v^R_{n+1}(y_1,\ldots, y_n, y_{n+1})
 v^R_n(y_1,\ldots, y_n, z_{n+1}) - 1\bigg)
 \d y_1\ldots \d y_{n+1} \d z_{n+1}
\end{split}
\end{equation}
As before, in the region where both $y_{n+1}$ and
$z_{n+1}$ are less than $R/2$, the expression inside the integral
is zero. Without any loss of generality, we may assume that $y_n+1>R/2$.
In that case, the expression above is bounded by
 \begin{equation}\nonumber
 \begin{split}
  (n+1) R^{1+\delta}\|\psi_{n+1}(y_1, \ldots, y_n, y_n+1)
   \chi(|y_n+1|\geq R/2) G(-y_n+1)\|^2\\
   +
  (n+1) R^{-(1+\delta)}\|\psi_{n+1}(y_1, \ldots, y_n, z_n+1)
  G(-z_n+1)\|^2
 \end{split}
 \end{equation}
Similarly to \eqref{eq:50}, we obtain
 $$
  \|\psi_{n+1}(y_1, \ldots, y_n, y_n+1)
   \chi(|y_n+1|\geq R/2) G(-y_n+1)\|^2 \leq R^{-2(1+\delta)}
   \|\psi_{n+1}\|^2 .
 $$
Therefore,
\begin{equation}\nonumber
\begin{split}
\langle D^*(0)D(0) \mathcal{U}^R\psi,
 \mathcal{U}^R\psi \rangle + \langle
 D^*(0)D(0) \mathcal{V}^R\psi, \mathcal{V}^R\psi
 \rangle - \langle D^*(0)D(0) \psi, \psi \rangle
 \leq \frac{c}{R^{1+\delta}} .
\end{split}
\end{equation}
This concludes the proof.
\end{proof}

\subsubsection{Localization error for the operator $\sigma .
B_f(0)$}
\begin{lemma}\label{lem6}
Let $\psi\in \mathfrak{Q}({\sigma.B_f(0)})\cap\mathfrak{D}(N_f)$,
and assume that there exists $p_0\in (6/5,2],$ such that
 $$
  \|\nabla_k a_\lambda(k) \psi\| \in L^{p_0}(\R^3) + L^2(\R^3).
 $$
Then, the inequality
\begin{equation}
 \langle \sigma.B_f(0) \mathcal{U}^R\psi, \mathcal{U}^R\psi
 \rangle + \langle
  \sigma.B_f(0)
 \mathcal{V}^R\psi, \mathcal{V}^R\psi \rangle - \langle
 \sigma.B_f(0) \psi, \psi
 \rangle
  \leq \frac{c}{R^{1+\delta}}
\end{equation}
holds with $\delta = (p_0 - 6/5)/2$.
\end{lemma}
\noindent The proof of Lemma~\ref{lem6} is similar to the one of
Lemma~\ref{lem5}, with a large number of simplifications.

\subsection{Proof of Theorem~\ref{thm:self-energy}}
We let
\begin{equation}
  \Phi^R: = \mathcal{U}^R \Omega_0,
\end{equation}
where $\Omega_0$ is a normalized ground state eigenfunction of the
operator $T(0)$, and where $\mathcal{U}^R$ is defined in \eqref{defu}.
We recall that we have $\langle T(0)\Omega_0, \Omega_0\rangle = \Sigma_0
\| \Omega_0 \| ^2$. We would like to show that the value of the
quadratic form associated to $T(0)$ at $\Phi^R$ is, for large
$R$, close to the value of the quadratic form associated to $T(0)$ at
$\Omega_0$.

First, we notice that $\Omega_0$ fulfills all the conditions of
Lemmata~\ref{lem1}-\ref{lem6} which implies that
 $$
  \langle T(0)\Omega_0,
  \Omega_0\rangle = \langle T(0)\mathcal{U}^R \Omega_0, \mathcal{U}^R
  \Omega_0\rangle + \langle T(0)\mathcal{V}^R \Omega_0, \mathcal{V}^R
  \Omega_0\rangle + \frac{C(R)}{R} ,
 $$
where $C(R)$ tends to zero as $R$ tends to infinity. Thus, since
$\langle T(0) \mathcal{V}^R \Omega_0, \mathcal{V}^R
\Omega_0\rangle\! \geq\! \Sigma_0 \| \mathcal{V}^R \Omega_0 \|^2$,
we obtain
\begin{eqnarray*}
 \langle T(0) \Phi^R, \Phi^R \rangle & \leq &
 \Sigma_0 + \frac{|C(R)|}{R} - \Sigma_0
 \langle T(0) \mathcal{V}^R \Omega_0 , \mathcal{V}^R \Omega_0\rangle
 \\ & \leq & \Sigma_0
 (1 - \|\mathcal{V}^R \Omega_0\|^2) + \frac{|C(R)|}{R}
 = \Sigma_0 \|\Phi^R \|^2 + \frac{|C(R)|}{R}\ ,
\end{eqnarray*}
which proves ii) of Theorem~\ref{thm:self-energy}.

To complete the proof of Theorem~\ref{thm:self-energy}, it
suffices to prove the two Inequalities~\eqref{eq:self-energy3} and
\eqref{eq:self-energy4}. Let us start with \eqref{eq:self-energy3}.
\begin{equation}
 \begin{split}
 | \langle D(x) \Phi^R, \Phi^R \rangle | &\leq \sum_n
 \sqrt{n+1}
 \int |G(x-y_{n+1})|~| \Phi^R_{n+1}|~| \Phi^R_n| \d y_1 \d y_{n+1}
 \\ &= \sum_n\! \sqrt{n+1}
 \frac{2}{|x|}\!\int_{|y_{n+1}| \leq R}
 \!\!\!|G(x-y_{n+1})|(1+|x-y_{n+1}|)
 |\Phi^R_{n+1}|\\
 &\hspace{4cm}\times\frac{(1+|y_{n+1}|)^{2\delta}}
 {(1+|y_{n+1}|)^{2 \delta}} |\Phi^R_n| \d y_1 \d y_{n+1} \nonumber
 \end{split}
\end{equation}
By applying the Schwarz inequality, we get
\begin{eqnarray}
 | \langle D(x) \Phi^R, \Phi^R \rangle | &\leq& \sum_n
 \sqrt{n+1}
 \int |G(x-y_{n+1})|~| \Phi^R_{n+1}|~| \Phi^R_n|
 \d y_1 \d y_{n+1}
  \label{eq:thm1-1}\\ & = & \sum_n\! \frac{2}{|x|} \|
 G(x-y_{n+1})(1+|x-y_{n+1}|) (1+|y_{n+1}|)^{-2\delta} \Phi^R_n \|
    \nonumber\\
 &&\hspace{3cm}\times \|
 \sqrt{n+1}(1+|y_{n+1}|)^{2\delta} \Phi^R_{n+1}\|\nonumber
\end{eqnarray}
We recall that from Lemma~\ref{lem-appendix1} that
$|G(x-y_{n+1})(1+|x-y_{n+1}|)|\in L^r(\R^3)$ for all $r>2$.
Therefore, for $p >3/(3-2\delta)$, and $q$ given by $1/p + 1/q=1$,
we have $\|(1+|y_{n+1}|)^{-2\delta}\|_q <\infty$. Thus,
\begin{eqnarray*}
\lefteqn{\| G(x-y_{n+1})(1+|x-y_{n+1}|)
 (1+|y_{n+1}|)^{-2\delta} \Phi^R_n \|} & & \\
 & \leq & \|G(x-y_{n+1})(1+|x-y_{n+1}|)\chi(|y_{n+1}|\leq R)\|_p
 \|(1+|y_{n+1}|)^{-2\delta}\|_q \|\Phi_n^R\|.
\end{eqnarray*}
Moreover,  for $|x|>2R$, the norm
$\|G(x-y_{n+1})(1+|x-y_{n+1}|)\chi(|y_{n+1}|\leq R)\|_p$ tends to
zero as $R\rightarrow\infty$. This estimate together with \eqref{eq:thm1-1} yields
\begin{eqnarray}
 | \langle D(x) \Phi^R, \Phi^R \rangle |   &\leq &
 \! \frac{2}{|x|} \varepsilon(x) \sum_n\left( \| \Phi^R_n \|^2
 + \| \sqrt{n+1}(1+|y_{n+1}|)^{2\delta}
 \Phi^R_{n+1}\|^2\right)\nonumber
\end{eqnarray}
Conditions $\mathfrak{C}_1 iii)$ and $\mathfrak{C}_1 iv)$ together
with the above inequality conclude the proof of
\eqref{eq:self-energy3} if we pick $\delta = (p_0-6/5)/2$. The
proofs of \eqref{eq:self-energy3.5}, \eqref{eq:20.5}, and
\eqref{eq:self-energy4} are similar.

\section{Approximate ground state for a system with an external
potential} 
In the present section, we consider the Pauli-Fierz
Hamiltonian for $M$ electrons with an external potential
\begin{equation}\nonumber
\begin{split}
H_{M} = & \sum_{\ell=1}^{M} \left\{\left(-i\nabla_{x_\ell}\otimes I_f +
\sqrt{\alpha} A_f(x_\ell)\right)^2 + \sqrt{\alpha}\sigma\cdot B_f(x_\ell)
+ V(x_\ell)\otimes I_f\right\}\\ & + \frac12 \sum_{1\leq k,\ell\leq M}W(x_k -
x_\ell)\otimes I_f + I_{el}\otimes H_f\ ,
\end{split}
\end{equation}
acting on $\mathcal{H} = \mathcal{H}_M^{el}\otimes\mathcal{F}$.
The brackets $\bra\,\cdot\, , \, \cdot\,\ket$ will from here on denote
the scalar product on ${\mathcal H}$. 
Furthermore, for the rest of this section, we will
write operators of the form $I_{el}\otimes A_f$ or  $B_{el}\otimes
I_f$ on ${\mathcal H}$ simply as $A_{f}$ or $B_{el}$,
respectively, in order not to overburden the notation. The precise
meaning will be clear from the context.

We assume that the Condition $\mathfrak{C}_2$ is fulfilled for
this system, which implies, in particular, that the operator
$H_{M}$ has a ground state. We will construct an approximation to
the ground state which is spatially localized with respect to the
electron and photon variables, and whose energy is close to the
ground state energy.

\subsection{Localization of the electrons}
We start with localization in the electron configuration space.
To this end, we recall from (~\ref{eq:HMgs}) that $\Upsilon$ denotes the
ground state of $H_M$.
For $u$ given by \eqref{def-smallu}, we define $\Upsilon^R =
(\Upsilon^R_0, \Upsilon^R_1, \ldots , \Upsilon^R_n, \ldots)\in
\mathcal{H} = \mathcal{H}_M^{el}\otimes\mathcal{F}$ by
 $$ \Upsilon^R_n =  u\left(\frac{2\sqrt{\sum_{i=1}^M
   |x_i|^2}}{R}\right)  \Upsilon_n\ ,
 $$
where $\Upsilon_n$ is the $n$-photon component of
$\Upsilon$.
Notice that on the support of $\Upsilon^R$, we have $|x_i|\le R/2$ for
$i=1, \ldots , M.$
\begin{lemma}\label{lem:x-loc}
For all $R > 1$,
\begin{equation}\label{eq:lem:x-loc1}
 \bra H_{M} \Upsilon^R, \Upsilon^R \ket \leq
 E_{M} + \frac{c}{R^2}
\end{equation}
\begin{equation}\label{eq:lem:x-loc2}
 1- \frac{c}{R^2} \leq \| \Upsilon^R \| \leq 1
\end{equation}
\end{lemma}
The proof of this Lemma follows immediately from standard
localization error estimates for Schr\"odinger operators
\cite{Cyconetal1987}, and the Condition~$\mathfrak{C_2}$ iii).

\subsection{Localization of photons}\label{S4.2}
Our next goal is to localize all photons in a ball of radius $2R$
centered at the origin. For this purpose, we define the function
$\Psi^R= (\Psi^R_1, \Psi^R_2, \ldots \Psi^R_n, \ldots )\in
\mathcal{H}_M^{el}\otimes\mathcal{F}$ as
\begin{equation}
\Psi^R = \mathcal{U}^{2R} \Upsilon^R .
\end{equation}
where $\mathcal{U}^R$ straightforwardly extends the operator
defined on $\gF$ in \eqref{defu} to
$\mathcal{H}_M^{el}\otimes\mathcal{F}$.

We note here that the localization radius for photons is chosen to
be four times larger than that for the electrons. The consequence
is that the contribution of the "external" photons to the magnetic
vector-potential will be negligible within the region where the
electrons are localized.

Similarly to Lemma~\ref{lem1}, we find that there exists $c<\infty$,
such that for all $\varepsilon>0$, and all $R$ large enough,
\begin{equation}\label{eq:loc-x-1}
\begin{split}
 \bra  H_f\mathcal{U}^{2R}\Upsilon^R, \mathcal{U}^{2R}
 \Upsilon^R \ket + \bra   H_f
 \mathcal{V}^{2R}\Upsilon^R, \mathcal{V}^{2R}\Upsilon^R \ket -
 \bra   H_f \Upsilon^R, \Upsilon^R \ket \\
   \leq  \bra
 N_f  \Upsilon^R, \Upsilon^R \ket \left( \frac{\varepsilon}{R}  +
 \frac{c}{\varepsilon R} \frac{\| \mathcal{V}^{R}
 \Upsilon^R\|^2}{\|\Upsilon^R\|^2} \right).
\end{split}
\end{equation}
Obviously, it suffices to compute the localization error only for
the operator
 $$
   (-i\nabla_{x_1}  + \sqrt{\alpha} A_f(x_1))^2 +
   \sqrt{\alpha} \sigma\cdot B_f(x_1) +   H_f.
 $$
In the rest of this section, we will denote $x=x_1$.

\begin{lemma}\label{lem-loc-pA}
The  following estimate holds
\begin{equation}\label{eq:lem-loc-pA0}
\begin{split}
  \left|\bra  \mathcal{U}^{2R}  \Upsilon^R, i\nabla_x A_f(x)
  \mathcal{U}^{2R} \Upsilon^R \ket +
  \bra \mathcal{V}^{2R} \Upsilon^R, i\nabla_x A_f(x)
  \mathcal{V}^{2R} \Upsilon^R \ket -
  \bra \Upsilon^R, i\nabla_x A_f(x) \Upsilon^R \ket \right| \\
  \leq  \frac{c}{R^{1+\delta}} \left( \|N_f
  \Upsilon^R \|^2 + \|\nabla_x
  \Upsilon^R \|^2\right),
\end{split}
\end{equation}
where $\delta = (p_0 - 6/5)/2$ and $p_0$ is given by
$\mathfrak{C}_2$~ii).
\end{lemma}
\begin{proof}
The proof of this Lemma is very similar to the one of
Lemma~\ref{lem4}.
\begin{eqnarray}\label{eq:lem-loc-pA1}
  \lefteqn{\ \hspace{-2cm}
  \left|\bra \mathcal{U}^{2R} \Upsilon^R, i\nabla_x D(x)
  \mathcal{U}^{2R} \Upsilon^R \ket +
  \bra \mathcal{V}^{2R} \Upsilon^R, i\nabla_x D(x)
  \mathcal{V}^{2R} \Upsilon^R \ket -
  \bra \Upsilon^R, i\nabla_x D(x) \Upsilon^R \ket \right|}
  \nonumber && \\
  & \leq &\int_{|x|\leq \frac{R}{2}} \d x \sum_n \sqrt{n+1} \int
  |G_\lambda(x-y_{n+1})|~|\Upsilon_{n+1}^R|~|\nabla_x
  \Upsilon_n^R|\\
  & & \times(u_n^{2R}u_{n+1}^{2R} + v_n^{2R} v_{n+1}^{2R} -1) \d y_1 \ldots
  \d y_{n+1} .\nonumber
\end{eqnarray}
Similarly to Lemma~\ref{lem4}, we show that $(u_n^{2R}u_{n+1}^{2R}
+ v_n^{2R} v_{n+1}^{2R} -1)$ is nonzero only if $|y_{n+1}| \geq
R$. This implies $ |x - y_{n+1}| \geq |y_{n+1}| /2 \geq R/2$.
Therefore, the integral in \eqref{eq:lem-loc-pA1} can be estimated
by
\begin{equation}
\begin{split}
 &\frac{1}{R^{1+\delta}} \int_{|x|\leq \frac{R}{2}} \d x
 \sum_n \sqrt{n+1} \|\Upsilon_{n+1}^R (1 +
 |y_{n+1} |)^{2\delta} \| \\
 &\ \ \ \times \|\sum_\lambda G_\lambda(x-y_{n+1})
 ( 1 + |x - y_{n+1}|)^{1-\delta}\| ~ \| \nabla_x \Upsilon_n^R\|
\end{split}
\end{equation}
Since the term $\| \nabla_x \Upsilon^R \|$ is finite, the rest of
the proof is not different from the one of Lemma~\ref{lem4}.
\end{proof}

Similarly to Lemmata~\ref{lem5} and \ref{lem6} and the above
Lemma~\ref{lem-loc-pA}, one can prove that
\begin{equation}\label{lem-loc-Asquare}
\begin{split}
 \left|\bra \mathcal{U}^{2R} \Upsilon^R,  A^2(x)
  \mathcal{U}^{2R} \Upsilon^R \ket\! +\!
  \bra \mathcal{V}^{2R} \Upsilon^R,  A^2(x)
  \mathcal{V}^{2R} \Upsilon^R \ket \!-\!
  \bra \Upsilon^R,  A^2(x) \Upsilon^R \ket \right|
  \!\leq\!  \frac{c}{ R^{1+\delta}}
\end{split}
\end{equation}
and
\begin{equation}\label{lem-loc-sigmaB}
\begin{split}
  \left|\bra \mathcal{U}^{2R}
  \Upsilon^R,  \sigma\cdot B_f(x)
  \mathcal{U}^{2R} \Upsilon^R \ket \!+\!
  \bra \mathcal{V}^{2R} \Upsilon^R,  \sigma\cdot B_f(x)
  \mathcal{V}^{2R} \Upsilon^R \ket \!-\!
  \bra \Upsilon^R,  \sigma\cdot B_f(x) \Upsilon^R \ket \right|
    \\
   \!\leq\!  \frac{c}{R^{1+\delta}}
\end{split}
\end{equation}

\begin{theorem}[Energy of the approximate ground
state]\label{thm:estimate2} For arbitrarily fixed $\varepsilon>0$
and $R$ large enough, the following statements hold.

\noindent i)
\begin{equation}\label{eq:thm:estimate2}
\begin{split}
 E_{M} \|\Psi^R\|^2\leq
 \bra H_{M} \Psi^R, \Psi^R \ket
 \leq  E_{M} \|\Psi^R\|^2 + \frac{\varepsilon}{R}
 \|\Psi^R\|^2
\end{split}
\end{equation}

\noindent ii) Let $z\in\R^3$ be an external variable, i.e., the
function $\Psi^R$ does not depend $on\ z$. Then, for $|z|>4R$
\begin{equation}\label{eq:67}
| \langle D(z) \Psi^R, \Psi^R \rangle  | \leq \frac{c(z)}{|z|}\ ,
\end{equation}
\begin{equation}\label{eq:68}
| \langle D(z)^2 \Psi^R, \Psi^R \rangle   | \leq \frac{c(z)}{|z|^2}\
,
\end{equation}
\begin{equation}\label{eq:68.5}
| \langle D^*(z)D(z) \Psi^R, \Psi^R \rangle   | \leq
\frac{c(z)}{|z|^2}\ ,
\end{equation}
and
\begin{equation}\label{eq:69}
| \langle K(z) \Psi^R, \Psi^R \rangle  | \leq \frac{c(z)}{|z|}\ ,
\end{equation}
where $c(z)$ is a function independent of $R$ that tends to zero
as $|z|$ tends to infinity.
\end{theorem}
\begin{proof}
Applying Lemma~\ref{lem:x-loc}, and
Inequalities~\eqref{eq:lem-loc-pA0}, \eqref{lem-loc-Asquare},
\eqref{lem-loc-sigmaB}, we obtain
\begin{equation}
\begin{split}
  \bra H_{M} \mathcal{U}^{2R}\Upsilon^R, \mathcal{U}^{2R}\Upsilon^R
  \ket + \bra H_{M} \mathcal{V}^{2R}\Upsilon^R,
  \mathcal{V}^{2R}\Upsilon^R \ket -\frac{\varepsilon}{R} \leq
  E_{M} + \frac{c}{R^2}
\end{split}
\end{equation}
Using $ E_{M} \| \mathcal{V}^{2R}\Upsilon^R \|^2 \leq \bra
H_{M} \mathcal{V}^{2R}\Upsilon^R, \mathcal{V}^{2R}\Upsilon^R
  \ket$, we get
\begin{equation}
\begin{split}
  \bra H_{M} \mathcal{U}^{2R}\Upsilon^R, \mathcal{U}^{2R}\Upsilon^R
  \ket \leq E_{M} \| \mathcal{U}^{2R}\Upsilon^R \|^2 +
  \frac{\varepsilon}{R} + \frac{c}{R^2}
\end{split}
\end{equation}
Since $\|\Upsilon^R\|\rightarrow 1$ as $R\rightarrow\infty$, we
get \eqref{eq:thm:estimate2}.

The proof of ii) is analogous to the proof of
Lemma~\ref{lem-loc-pA} and Theorem~\ref{thm:self-energy} iii).
\end{proof}

\section{Proof of Theorem~\ref{mainthm}}
The previous discussion enables us to construct a normalized trial
function $\Gamma^{R,b}\in\mathcal{H}_N^{el}\otimes\mathcal{F}$.
For given $N\in\N$, we define $\tilde\Psi^{R,N-1}$ as
 $$
  \tilde\Psi^{R,N-1} = \frac{\Psi^R}{\|\Psi^R\|},
 $$
where $\Psi^R$ is the function defined in Section~\ref{S4.2}
for a system of $M = N-1$ electrons.
As a natural candidate for a trial state for the proof of
Theorem~\ref{mainthm}, one could consider the state $\varphi =
(\varphi_0, \varphi_1, \ldots ) $ defined by
\begin{equation}
\begin{split}
  \varphi_n = \sum\limits_{j=0}^{n} &\Theta_j^{R,b}
  (y_1, \ldots , y_j, \lambda_1,
  \ldots ,   \lambda_j,        x_N, s_N)\cdot  \\
  &\times\tilde\Psi _{n-j}^{R, N-1}(y_{j+1}, \ldots , y_n,
  \lambda_{j+1}, \ldots , \lambda_n, x_1, \ldots , x_{N-1}, s_1,
  \ldots , s_{N-1}).
\end{split}
\end{equation}
However, since the components $\varphi_n$ are neither symmetric in
the photon, nor antisymmetric in the electron variables, our next
goal is to symmetrize the function $\varphi_n$ in the photon
variables, and to antisymmetrize it in the electron variables.

We denote by $S_{n, j}$ the set of $\cnj$ possible partitions $g$
of the set of $n$ indices $\{1, \ldots , n  \}$ into two subsets
$C_1$ and $C_2$ with $j$ and $n-j$ elements respectively. Let
$i_1(g),\ldots , i_j(g)$ be the indices in $C_1$ and $i_{j+1}(g),
\ldots , i_n(g)$ in $C_2.$ We define the function
\begin{equation}\nonumber
\begin{split}
    (\Pi_{n,j}^p(g)\Theta_j^{R, b}&\tilde\psi_{n-j}^{R, N-1})
        (y_1, \ldots , y_n, \lambda_1, \ldots , \lambda_n, x, s)\\
   :& = \Theta_j^{R, b}
        (y_{i_1}, \ldots , y_{i_j}, \lambda_{i_1}, \ldots , \lambda_{i_j}, x_N, s_N)\\
    &\times \tilde\psi_{n-j}^{R, N-1}
        (y_{i_{j+1}}, \ldots , y_{i_n}, \lambda_{i_{j+1}}, \ldots , \lambda_{i_n}, x_1,\ldots ,x_{N-1}, s_1, \ldots s_{N-1}) \;.
\end{split}
\end{equation}
Evidently,
\begin{equation}\label{deftildegamma}
   \tilde\Gamma_n^{R,b} := \sum_{j=0}^n \cnj^{-1/2}
   \sum_{g\in S_{n,j}}
   \Pi_{n,j}^p(g) \Theta_j^{R,b} \tilde\Psi_{n-j}^{R, N-1}\
\end{equation}
is symmetric with respect to the permutation of photon variables.

To construct a combination of the functions $\tilde\Gamma_n^{R,b}$
which is antisymmetric in the electron variables, let us consider the set of all transpositions $\pi_i$  $i= 1, \ldots , N$, which exchange a pair of electron variables $(x_i, s_i)$ with $(x_N, s_N)$, including the trivial transposition $(x_N, s_N) \leftrightarrow (x_N, s_N).$ For an arbitrary function $\varphi(x_1,\ldots ,x_N, s_1, \ldots , s_N)$, let
\begin{equation}\nonumber
(\Pi_i^{el}\varphi)(x_1,\ldots ,x_N, s_1, \ldots , s_N): = \varphi(\pi_i(x_1,\ldots ,x_N, s_1, \ldots , s_N)).
\end{equation}
Then, we define
\begin{equation}\label{def-trial}
\Gamma_n^{R,b} = \sum_{j=0}^n N^{-1/2} \cnj^{-1/2}\sum_{i=1}^N
\sum_{g\in S_{n,j}} (-1)^{\kappa(i)} \Pi_i^{el}
\Pi_{n,j}^p(g) \Theta_j^{R,b} \tilde\Psi_{n-j}^{R, N-1} ,
\end{equation}
where $\kappa(i)=0$ if $i=N$, and $\kappa(i)=1$ otherwise.
Obviously,
$$
\Gamma^{R,b}=(\Gamma_0^{R,b}, \Gamma_1^{R,b},
\ldots, \Gamma_n^{R,b}, \ldots) \in \mathcal{H}_N^{el} \otimes
\mathcal{F}.
$$
Notice that $\Gamma^{R,b}$ is a normalized function in
$\mathcal{H}_N^{el}\otimes\mathcal{F}$, since if $|b|>5 R$, the
summands
in \eqref{def-trial} have for different $i$ disjoint supports in electron variables, and thus
\begin{equation}\label{eq:normalization}
 \|\Gamma^{R,b}\|^2  = \|\tilde\Gamma^{R,b}\|^2= 1
\end{equation}
and
$$
(H_N \Gamma^{R, b},\ \Gamma^{R, b}) = (H_N \tilde\Gamma^{R, b},\ \tilde\Gamma^{R, b}).
$$
Although the state $\tilde\Gamma^{R, b}$ is not antisymmetric in all electron variables, the quadratic form of $H_N$ at $\tilde\Gamma^{R, b}$ is well-defined.

Furthermore, both functions $\Theta^{R,b}$ and $\tilde\Psi^{R,
N-1}$ have a finite expectation number of photons, say, $N_1$ and $N_2$,
respectively. Evidently, this implies that
$\Gamma^{R,b}$ has a finite expected photon number $N_1 +
N_2$.

We remark that for $|b|> 5R$, and each of the terms in the sum $$
\Pi_i^{el}\sum_{g\in S_{n,j}}\Pi_{n, j}^p(g)\Theta_j^{R,
b}\tilde\psi_{n-j}^{R, N-1} , $$ $\Theta_j^{R, b}$ and
$\tilde\psi_{n-j}^{R, N-1}$ have disjoint supports, thus one finds
 $$ (H_N \Gamma^{R, b},\ \Gamma^{R, b}) = (H_N \tilde\Gamma^{R,
   b},\ \tilde\Gamma^{R, b}),
 $$
where, as we recall from (~\ref{deftildegamma}), $\tilde\Gamma^{R,
b}= ( \tilde\Gamma_0^{R, b}, \tilde\Gamma_1^{R, b}, \ldots )$ is
the state prior to antisymmetrization in the electron variables.
Hence, instead of estimating the quadratic form of the operator
$H_N$ with respect to the state $\Gamma^{R, b}$, we may estimate
it with respect to $\tilde\Gamma^{R, b}.$ Although this state is
not antisymmetric in all electron variables, the quadratic form is
well-defined.

We recall that in our notation for the state $\tilde\Gamma^{R,
b}$, the variables $(x_N, s_N)$ are the arguments of $\Theta^{R,
b}$, while $(x_1, \ldots , x_{N-1}, s_1, \ldots , s_{N-1})$ are
the arguments of $\tilde\psi^{R, N-1}$, and furthermore, that
$\|\tilde\Gamma^{R, b} \| = 1.$

\begin{lemma}\label{lem:estimate4}
For $|b| >8R$, there exists $c>0$ independent of $R$ such
that the following estimate holds
\begin{equation}\nonumber
 \bra H_f \Gamma^{R,b}, \Gamma^{R,b}  \ket\! \leq\!
 \bra H_f \Theta^{R,b}, \Theta^{R,b} \ket +
 \bra H_f \tilde\Psi^{R, N-1}, \tilde\Psi^{R, N-1}\ket
 +  c\frac{R^{3/2}}{|b|^{5/2}}
 \bra N_f \Gamma^{R,b}, \Gamma^{R,b}\ket
\end{equation}
\end{lemma}
\begin{proof}
We have
\begin{equation}\label{eq:lem:estimate4-1}
 H_f \tilde\Gamma^{R,b}_n = n |\nabla_{y_1}|
  \sum_{j=0}^n \cnj^{-1/2}
   \sum_{g\in S_{n,j}}
   \Pi_{n,j}^p(g) \Theta_j^{R,b} \tilde\Psi_{n-j}^{R, N-1}\ .
\end{equation}
Let us start with one of the functions in the sum
\eqref{eq:lem:estimate4-1}. We take for example the expression
$n|\nabla_{y_1}|
\Theta^{R,b}_j (y_1, \ldots, y_j) \tilde\Psi^{R, N-1}_{n-j} (y_{j+1},
\ldots, y_n)$. All other terms can be treated similarly.

In the quadratic form $\langle H_f \tilde\Gamma^{R,b},
\tilde\Gamma^{R,b} \rangle$, this term appears twice, in
\begin{equation}
 \begin{split}
 n\Bra |\nabla_{y_1}| &\Theta^{R,b}_j (y_1,
 \ldots, y_j) \tilde\Psi^{R, N-1}_{n-j} (y_{j+1}, \ldots, y_n),
 \\
 &\Theta^{R,b}_j (y_1, \ldots, y_j) \tilde\Psi^{R, N-1}_{n-j} (y_{j+1},
 \ldots, y_n) \Ket , \nonumber
 \end{split}
\end{equation}
and in
\begin{equation}\label{eq:72}
  \begin{split}
  n\Bra |\nabla_{y_1}| &\Theta^{R,b}_j (y_1,
 \ldots, y_j) \tilde\Psi^{R, N-1}_{n-j} (y_{j+1}, \ldots, y_n),
 \\
 &\Theta^{R,b}_{j-1} (y_2, \ldots, y_j) \tilde\Psi^{R, N-1}_{n-j+1} (y_1, y_{j+1},
 \ldots, y_n) \Ket .
 \end{split}
\end{equation}
All other cross terms appearing in the quadratic form $\langle H_f
\tilde\Gamma^{R,b}, \tilde\Gamma^{R,b}\rangle$ that contain the
function $n|\nabla_{y_1}| \Theta^{R,b}_j (y_1, \ldots, y_j)
\tilde\Psi^{R, N-1}_{n-j} (y_{j+1}, \ldots, y_n)$ are zero,
because at least for one variable, the supports of the functions
in the scalar product are disjoint. Let us now estimate
\eqref{eq:72}. The function
 $$
  \Theta^{R,b}_j (y_1, \ldots, y_j)
  \tilde\Psi^{R, N-1}_{n-j} (y_{j+1}, \ldots, y_n)
 $$
is supported in
the region $\{ |y_1| \geq |b| -2R \}$ whereas
 $$
  \Theta^{R,b}_{j-1}
  (y_2, \ldots, y_j) \tilde\Psi^{R, N-1}_{n-j+1} (y_1, y_{j+1},
  \ldots, y_n)
 $$
is supported in the region $\{ |y_1| \leq 2R \}$. Applying
Lemma~\ref{lem:loc-estimate-modulusp} with $|b|> 8 R$, we arrive
at
\begin{equation}
\begin{split}
 n\Big|\Bra |\nabla_{y_1}| &\Theta^{R,b}_j (y_1,
 \ldots, y_j) \tilde\Psi^{R, N-1}_{n-j} (y_{j+1}, \ldots, y_n), \\
 &\Theta^{R,b}_{j-1} (y_2, \ldots, y_j) \tilde\Psi^{R, N-1}_{n-j+1} (y_1, y_{j+1},
 \ldots, y_n) \Ket\Big| \\
 \leq c~n\frac{R^{3/2}}{|b|^{5/2}}
 \Big(\|  &\Theta^{R,b}_j (y_1,
 \ldots, y_j) \tilde\Psi^{R, N-1}_{n-j} (y_{j+1}, \ldots, y_n) \|^2 \\
 + \| &\Theta^{R,b}_{j-1} (y_2, \ldots, y_j) \tilde\Psi^{R, N-1}_{n-j+1}
  (y_1, y_{j+1}, \ldots, y_n)  \|^2\Big) ,
\end{split}
\end{equation}
which implies
\begin{equation}\label{eq:74}
\begin{split}
 \lefteqn{\bra H_f \tilde\Gamma^{R,b}, \tilde\Gamma^{R,b} \ket
 \leq   c \frac{R^{3/2}}{|b|^{5/2}}\langle N_f \tilde\Gamma^{R,b},
 \tilde\Gamma^{R,b}\rangle} &\\
 & + \sum_n n \sum_{j=0}^n \cnj^{-1}
   \sum_{g\in S_{n,j}}
   \bra |\nabla_{y_1}| \Pi_{n,j}^p(g) \Theta_j^{R,b}
   \tilde\Psi_{n-j}^{R, N-1}, \Pi_{n,j}^p(g) \Theta_j^{R,b}
   \tilde\Psi_{n-j}^{R, N-1}\ket .
\end{split}
\end{equation}
For fixed $n$ and $j$, in the sum
 $$ \sum_{g\in S_{n,j}}
  \bra |\nabla_{y_1}| \Pi_{n,j}^p(g) \Theta_j^{R,b}
  \tilde\Psi_{n-j}^{R, N-1}, \Pi_{n,j}^p(g) \Theta_j^{R,b}
  \tilde\Psi_{n-j}^{R, N-1}\ket,
 $$
the variable $y_1$ appears
$n-1\choose j-1$ times in $\Theta^{R,b}_j$ and $n-1\choose n-j-1$
times in $\tilde\Psi^{R, N-1}_{n-j}$. Therefore, the second term
on the right hand side of \eqref{eq:74} can be rewritten as
\begin{equation}\label{eq:75}
 \begin{split}
 \sum_n &\sum_{j=1}^n n \cnj^{-1} {n-1 \choose j-1} \langle |\nabla_{y_1}|
 \Theta^{R,b}_j(y_1, \ldots, y_j), \Theta^{R,b}_j(y_1, \ldots, y_j)
 \rangle \|\tilde\Psi^{R, N-1}_{n-j}\|^2 \\
 &+ \sum_n \sum_{j=1}^n n \cnj^{-1} {n-1 \choose n-j-1}
 \|\Theta^{R,b}_j\|^2\\
 &\hspace{2cm}\langle |\nabla_{y_1}| \tilde\Psi^{R,
 N-1}_{n-j}(y_1, \ldots y_{n-j}), \tilde\Psi^{R,
 N-1}_{n-j}(y_1, \ldots y_{n-j}) \rangle \\
 =& \sum_n \sum_{j=1}^n j \langle |\nabla_{y_1}|
 \Theta^{R,b}_j(y_1, \ldots, y_j), \Theta^{R,b}_j(y_1, \ldots, y_j)
 \rangle \|\tilde\Psi^{R, N-1}_{n-j}\|^2 \\
 &+ \sum_n \sum_{j=1}^n (n-j)
 \|\Theta^{R,b}_j\|^2 \langle |\nabla_{y_1}| \tilde\Psi^{R,
 N-1}_{n-j}(y_1, \ldots y_{n-j}), \tilde\Psi^{R,
 N-1}_{n-j}(y_1, \ldots y_{n-j}) \rangle \\
 =& \langle H_f \Theta^{R,b}, \Theta^{R,b}\rangle
 + \langle H_f \tilde\Psi^{R, N-1}, \tilde\Psi^{R, N-1} \rangle
 \end{split}
\end{equation}
The relations \eqref{eq:74} and \eqref{eq:75} imply the statement of
the Lemma.
\end{proof}

\begin{lemma}
For any $\varepsilon>0$ and $|b|$ large enough,
\begin{equation}\label{eq:80}
\begin{split}
  \lefteqn{\langle \sum_{\ell=1}^{N} i\nabla_{x_\ell}A(x_\ell)\tilde\Gamma^{R,b},
  \tilde\Gamma^{R,b}\rangle}  &\\
  & \leq \langle
  i\nabla_{x_N}A(x_N)\Theta^{R,b},
  \Theta^{R,b}\rangle +
  \sum_{\ell=1}^{N-1} \langle i\nabla_{x_\ell}A(x_\ell)\tilde\Psi^{R,N-1},
  \tilde\Psi^{R, N-1}\rangle\\
  & \ \ + \frac{\varepsilon}{2(|b|-2R)} \left( \| \nabla_{x_N}
  \Theta^{R,b} \|^2 + \| \Theta^{R,b} \|^2 \right)\\
  & \ \ + \sum_{\ell=1}^{N-1}
  \frac{\varepsilon}{2(|b|-2R)} \left( \| \nabla_{x_\ell}
  \tilde\Psi^{R,N-1} \|^2 + \| \tilde\Psi^{R,N-1} \|^2 \right) .
\end{split}
\end{equation}
Furthermore,
\begin{equation}\label{eq:82}
\begin{split}
  \lefteqn{\langle \sum_{\ell=1}^{N} \sigma\cdot B(x_\ell)\tilde\Gamma^{R,b},
  \tilde\Gamma^{R,b}\rangle} & \\
  & \leq \langle
  \sigma\cdot B(x_N)\Theta^{R,b},
  \Theta^{R,b}\rangle +
  \sum_{\ell=1}^{N-1} \langle \sigma\cdot B(x_\ell)\tilde\Psi^{R,N-1},
  \tilde\Psi^{R, N-1}\rangle\\
  & \ \ + \frac{\varepsilon}{(|b|-2R)}
  \|\Theta^{R,b} \|^2
  + \sum_{\ell=1}^{N-1}
  \frac{\varepsilon}{(|b|-2R)}
  \|\tilde\Psi^{R,N-1} \|^2\ ,
\end{split}
\end{equation}
and
\begin{equation}\label{eq:83}
\begin{split}
\left|\langle \sum_{\ell=1}^{N}\! D^2(x_\ell)\tilde\Gamma^{R,b},
  \tilde\Gamma^{R,b}\rangle\! -\! \langle
  D^2(x_N)\Theta^{R,b},
  \Theta^{R,b}\rangle \! -\!\!
  \sum_{\ell=1}^{N-1} \langle D^2(x_\ell)\tilde\Psi^{R,N-1},
  \tilde\Psi^{R, N-1}\rangle\right|\\
   \leq \frac{\varepsilon}{(|b|-2R)}
  \|\Theta^{R,b} \|^2
  + \sum_{\ell=1}^{N-1}
  \frac{\varepsilon}{(|b|-2R)}
  \|\tilde\Psi^{R,N-1} \|^2
\end{split}
\end{equation}
Moreover,
\begin{equation}\label{eq:84}
\begin{split}
\bigg|\langle \sum_{\ell=1}^{N}\!
D^*(x_\ell)D(x_\ell)\tilde\Gamma^{R,b},
  \tilde\Gamma^{R,b}\rangle\! -\! \langle
  D^*(x_N)D(x_N)\Theta^{R,b},
  \Theta^{R,b}\rangle \\ -
  \sum_{\ell=1}^{N-1} \langle D^*(x_\ell)
  D(x_\ell)\tilde\Psi^{R,N-1},
  \tilde\Psi^{R, N-1}\rangle\bigg|\\
   \leq \frac{\varepsilon}{(|b|-2R)}
  \|\Theta^{R,b} \|^2
  + \sum_{\ell=1}^{N-1}
  \frac{\varepsilon}{(|b|-2R)}
  \|\tilde\Psi^{R,N-1} \|^2
\end{split}
\end{equation}
\end{lemma}
\begin{proof}

We recall that in $\Theta_j^{R,b} \tilde\Psi_{n-j}^{R, N-1}$, the
variable $x_1$ appears only in the function $\tilde\Psi_{n-j}^{R,
N-1}$, and the variable $x_N$ only in $\Theta^{R,b}_j$. Permutations of
photon variables do not change this fact. We have, for
$k=1,\ldots, N$,
\begin{equation}\label{eq:86}
 \begin{split}
 &\left(D(x_k)\tilde\Gamma^{R,b}\right)_{n-1} \\
 &= \sum_{j=0}^n \cnj^{-1/2}
 \sum_{g\in S_{n,j}}\sqrt{n}
 \langle G(x_k - y_n),
 \Pi_{n,j}^p(g)  \Theta_j^{R,b}
 \tilde\Psi_{n-j}^{R, N-1} \rangle_{L^2(\R^3\otimes\C^2, \d y_n )}\  ,
 \end{split}
\end{equation}
where as before, $\d y_n$ means integration with respect to $y_n$
and summation over the associated polarization $\lambda_n$.

Let us start with one of the functions
$\Pi_{n,j}^p(g)\Theta_j^{R,b}\tilde\Psi_{n-j}^{R, N-1}$ in the sum
\eqref{deftildegamma}. For fixed $g$, two variants are possible.
Either the index $n$ is in $C_1$, and the function
$\Theta_j^{R,b}$ depends on the photon variable $y_n$, or the
function $\tilde\Psi_{n-j}^{R, N-1}$ depends on $y_n$. For fixed
$n$ and $j$, the first variant occurs $n-1\choose j-1$ times,
whereas the second one occurs $n-1\choose n-j-1$ times. Let us
consider the function
 $$
  \cnj^{-1/2}\Theta_j^{R,b}(x_N, y_1, \ldots y_j)
  \tilde\Psi_{n-j}^{R, N-1}(x_1, \ldots x_{N-1}, y_{j+1}, \ldots
  y_{n})
 $$
In the quadratic form $\langle\sum_{k=1}^N i\nabla_{x_k}
D(x_k)\tilde{\Gamma}^{R,b}, \tilde{\Gamma}^{R,b}\rangle$, it
appears only once in the scalar product with
 \begin{equation}
 \begin{split}
  \sqrt{n}\nabla_{x_k}\overline{G(x_k-y_n)}{n-1\choose j-1}^{-1/2}
  &\Theta_j^{R,b}(x_N, y_1, \ldots y_j)\\
  \times &\tilde\Psi_{n-j}^{R, N-1}(x_1, \ldots x_{N-1}, y_{j+1}, \ldots
  y_{n-1})\nonumber
 \end{split}
 \end{equation}
which, in the case $k\neq N$,  is equal to
\begin{equation}
\begin{split}
 \sqrt{n-j}&{n-1\choose n-j-1}^{-1}
 \Big\langle \tilde\Psi_{n-j}^{R, N-1}(x_1, \ldots x_{N-1}, y_{j+1}, \ldots
  y_{n}),  \\
  &\nabla_{x_k}\overline{G(x_k-y_n)} \tilde\Psi_{n-j-1}^{R, N-1}
  (x_1, \ldots x_{N-1}, y_{j+1}, \ldots
  y_{n-1})\Big\rangle \|\Theta_j^{R,b}\|^2\ ,
\end{split}
\end{equation}
and in the case $k=N$,
\begin{equation}
\begin{split}
 \sqrt{n-j}&{n-1\choose n-j-1}^{-1} \\
 &\Big\langle \tilde\Psi_{n-j}^{R, N-1}(x_1, \ldots x_{N-1}, y_{j+1}, \ldots
  y_{n})\nabla_{x_N}\Theta_j^{R,b}(x_N, y_1,\ldots y_j),  \\
  &\overline{G(x_N-y_n)} \tilde\Psi_{n-j-1}^{R, N-1}
  (x_1, \ldots x_{N-1}, y_{j+1}, \ldots
  y_{n-1})\Theta_j^{R,b}
  (x_N, y_1,\ldots y_j)\Big\rangle
\end{split}
\end{equation}
All other terms in \eqref{deftildegamma} with the same $j$, and
with $y_n$ in $\tilde\Psi_{n-j}^{R, N-1}$, give the same
contribution to $\langle\sum_{k=1}^N i\nabla_{x_k}
D(x_k)\tilde{\Gamma}^{R,b}, \tilde{\Gamma}^{R,b}\rangle$. Summing
up these $n-1\choose n-j-1$ contributions in the case $k\neq N$
yields
\begin{equation}\label{eq:89}
\begin{split}
 \sqrt{n-j}
 \Big\langle \tilde\Psi_{n-j}^{R, N-1},
  \nabla_{x_k}\overline{G(x_k-y_n)} \tilde\Psi_{n-j-1}^{R, N-1}
  \Big\rangle \|\Theta_j^{R,b}\|^2\ ,
\end{split}
\end{equation}
and in the case $k=N$,
\begin{equation}\label{eq:90}
\begin{split}
 \sqrt{n-j}
 \Big\langle \tilde\Psi_{n-j}^{R, N-1} \nabla_{x_N}\Theta^{R,b}_j,
  \overline{G(x_N-y_n)} \tilde\Psi_{n-j-1}^{R, N-1} \Theta^{R,b}_j
  \Big\rangle \ .
\end{split}
\end{equation}
If we sum first over $m=n-j$, and then  the terms
\eqref{eq:89} over $j$, we get
\begin{equation}\label{eq:91}
\langle \tilde\Psi^{R, N-1}, \nabla_{x_k}D(x_k) \tilde\Psi^{R,
N-1}\rangle \|\Theta ^{R,b}\|^2 .
\end{equation}
Let us compute first the sum over $n-j$ of the terms \eqref{eq:90},
and estimate them according to \eqref{eq:67}. We obtain for
$\varepsilon>0$, and $|b|$ sufficiently large,
\begin{equation}
\langle |\nabla_{x_N} \Theta_j^{R,b}|, \frac{c(x_N)}{|x_N|}
|\Theta_j^{R,b}| \rangle \leq \frac{\varepsilon}{2(|b|-2R)} \left(
\| \nabla_{x_N} \Theta_j^{R,b} \|^2 + \| \Theta_j^{R,b} \|^2
\right) ,
\end{equation}
where we used that $|x_N| \geq |b|-2R$, and $c(x_N)$ tends to zero,
as $|x_N|$ tends to infinity. Therefore,
\begin{equation}\label{eq:93}
\begin{split}
 \left|\sum_n\sum_j \sqrt{n-j}
 \Big\langle \tilde\Psi_{n-j}^{R, N-1} \nabla_{x_N}\Theta^{R,b}_j,
  \overline{G(x_N-y_n)} \tilde\Psi_{n-j-1}^{R, N-1} \Theta^{R,b}_j
  \Big\rangle\right| \\
 \leq \frac{\varepsilon}{2(|b|-2R)} \left( \| \nabla_{x_N}
\Theta^{R,b} \|^2 + \| \Theta^{R,b} \|^2 \right)
\end{split}
\end{equation}
In analogy to \eqref{eq:91} and \eqref{eq:93}, the contribution to
$\langle\sum_{k=1}^N i\nabla_{x_k} D(x_k)\tilde{\Gamma}^{R,b},
\tilde{\Gamma}^{R,b}\rangle$ of the terms for which the variable
$y_n$ is in $\Theta^{R,b}$, is, for $k=N$, equal to
\begin{equation}
\langle i \nabla_{x_N} D(x_N) \Theta^{R,b}, \Theta^{R,b}\rangle
\end{equation}
and for $k\neq N$, it can be estimated by
\begin{equation}
\frac{\varepsilon}{2(|b|-2R)} \left( \| \nabla_{x_k}
\tilde\Psi^{R,N-1} \|^2 + \| \tilde\Psi^{R,N-1} \|^2 \right)
\end{equation}
This completes the proof of \eqref{eq:80}.

\noindent Let us next prove the inequality \eqref{eq:83}.
The operator $D^2(x_k)$ acts as
\begin{equation}\label{eq:96}
\begin{split}
 \lefteqn{\left(D^2(x_k)\tilde\Gamma^{R,b}\right)_{n-2} = \sum_{j=0}^n \cnj^{-1/2}
 \sum_{g\in S_{n,j}}\sqrt{n}\sqrt{n-1}} &\\
 & \times\langle G(x_k - y_n)G(x_k - y_{n-1}), \Pi_{n,j}^p(g)\Theta_j^{R,b}
 \tilde\Psi_{n-j}^{R, N-1} \rangle_{L^2(X, \d y_n )\otimes
 L^2(X, \d y_{n-1} )}\ ,
\end{split}
\end{equation}
where $X:=\R^3\otimes\C^2$.
Assume that in the decomposition $g$, we have the
indices $n\in C_2$ and $(n-1)\in C_2$.
Then, both variables $y_n$ and $y_{n-1}$ appear in the function
$\tilde\Psi^{R, N-1}_{n-j}$. For fixed $n$ and $j$, we have
$n-2\choose n-j-2$ such cases. Similar to \eqref{eq:89} in the
case $k\neq N$, and to \eqref{eq:90} in the case $k=N$, we obtain,
respectively,
\begin{equation}
  \sqrt{n-j}\sqrt{n-1-j}  \Big\langle \tilde\Psi_{n-j}^{R, N-1},
  \overline{G(x_k-y_n)}~\overline{G(x_k - y_{n-1})}
  \tilde\Psi_{n-j-2}^{R, N-1}
  \Big\rangle \|\Theta_j^{R,b}\|^2\ ,
\end{equation}
and
\begin{equation}
\begin{split}
 \sqrt{n-j}\sqrt{n-j-1}
 \Big\langle \tilde\Psi_{n-j}^{R, N-1} \Theta^{R,b}_j,
  \overline{G(x_N-y_n)}~\overline{G(x_N - y_{n-1})}
  \tilde\Psi_{n-j-1}^{R, N-1} \Theta^{R,b}_j
  \Big\rangle \ .
\end{split}
\end{equation}
Now, summing each of these expressions over $m=n-j$ and $j$, and
applying \eqref{eq:68}, we arrive at
\begin{equation}\label{eq:99}
 \langle D^2(x_k)\tilde\Psi^{R, N-1}, \tilde\Psi^{R, N-1}\rangle
 \|\Theta^{R,b}\|^2
\end{equation}
for $k\neq N$, and
\begin{equation}\label{eq100}
 \frac{\varepsilon}{(|b| -2R)^2} \| \Theta^{R,b}\|^2
\end{equation}
for $k=N$.

Let us now consider $g$ with $n\in C_1$ and $(n-1)\in C_1$, which
implies that the variables $y_n$
and $y_{n-1}$ are in $\Theta^{R,b}_j$. We get
\begin{equation}\label{eq:101}
 \langle D^2(x_N)\Theta^{R, b}, \Theta^{R, b}\rangle
 \|\tilde\Psi^{R, N-1}\|^2
\end{equation}
for $k = N$, and
\begin{equation}\label{eq:102}
 \frac{\varepsilon}{(|b| -2R)^2} \| \tilde\Psi^{R, N-1}\|^2
\end{equation}
for $k \neq N$.

Finally, let us address the case where one of the indices $n$,
$n-1$ belongs to $C_1$ and the other one to $C_2.$
In this case, one of the variables $y_n$ and $y_{n-1}$ appears in
$\tilde\Psi^{R, N-1}_{n-j}$, and the other one in
$\Theta^{R,b}_j$. We have $2{n-2\choose j-1}$ such cases. Note
that in each such case, either $|G(x_k - y_n)|$ or $|G(x_k - y_{n-1})|$ is
small, and the contribution of the sum of these terms can be
estimated as
\begin{eqnarray}\label{eq:103}
 \frac{\varepsilon_1}{|b|-2R} \left( \langle N_f \Theta^{R,b},
 \Theta^{R, b}\rangle + \langle N_f \tilde\Psi^{R, N-1},
 \tilde\Psi^{R, N-1} \rangle \right)\\
 \leq
 \frac{\varepsilon}{|b|-2R} \left( \|\Theta^{R, b}\|^2 +
 \|\tilde\Psi^{R, N-1}\|^2 \right)\ .\nonumber
\end{eqnarray}
The estimates \eqref{eq:99}-\eqref{eq:103} imply \eqref{eq:83}

\noindent The proof of \eqref{eq:82} is very similar to
the one of \eqref{eq:80}.
\end{proof}

\subsection{Proof}
To prove Theorem~\ref{mainthm}, we will show that for suitably
chosen parameters $R$ and $|b|$, the trial function $\Gamma^{R,b}$
satisfies
\begin{equation}\label{eq:try}
 \langle H_{N} \Gamma^{R,b}, \Gamma^{R, b}\rangle
 < E_{N-1} + \Sigma_0  .
\end{equation}
We recall that
\begin{eqnarray}
 H_N  = \!\sum_{\ell=1}^{N} \left\{\left(-i\nabla_{x_\ell}
 + \sqrt{\alpha} A_f(x_\ell)\right)^2 + \sqrt{\alpha}\sigma\cdot B_f(x_\ell)
 + V(x_\ell) \right\}  \nonumber\\
 +\! \frac12\sum_{1\leq k,\ell\leq N}
 W(|x_k - x_\ell|) + H_f\ ,
\end{eqnarray}
and that, as was shown in the previous section, the
inequality~\eqref{eq:try} is equivalent to
\begin{equation}\nonumber
 \langle H_{N} \tilde\Gamma^{R,b}, \tilde\Gamma^{R, b}\rangle
 < E_{N-1} + \Sigma_0  .
\end{equation}
For $M\in\N$, we define
 $$
  I_M(x_1,\ldots, x_M) = \sum_{\ell = 1}^M V(x_\ell) +
  \frac12 \sum_{1\leq k,\ell\leq M}W(x_k-x_\ell).
 $$
Obviously, we have
\begin{equation}\label{eq:proof1}
\begin{split}
 \lefteqn{\sum_{\ell=1}^N \langle -\Delta_\ell \tilde\Gamma^{R,b},
 \tilde\Gamma^{R,b}\rangle + \langle I_N(x_1,\ldots, x_n)\tilde\Gamma^{R,b},
 \tilde\Gamma^{R,b}\rangle} & \\
 & = \sum_{\ell = 1}^{N-1} \langle -\Delta_\ell
 \tilde\Psi^{R, N-1}, \tilde\Psi^{R, N-1}\rangle\\
 & + \langle I_{N-1}(x_1, \ldots , x_{N-1}) \tilde\Psi^{R, N-1}
 , \tilde\Psi^{R, N-1}\rangle + \langle -\Delta
 \Theta^{R, b}, \Theta^{R, b}\rangle\\
 & + \left\langle V(x_N) + \langle \sum_{i=1}^{N-1} W(x_i-x_N)
 \tilde\Psi^{R,N-1}, \tilde\Psi^{R,N-1}\rangle \Theta^{R,b},
 \Theta^{R,b} \right\rangle
\end{split}
\end{equation}
where we used that $\Theta^{R, b}$ and $\tilde\Psi^{R, N-1}$ are
normalized. On the support of the function $\Theta^{R,b}$, we have
$|x_N|\leq |b| + R$ and on the support of the function
$\tilde\Psi^{R, N-1}$, $|x_i - x_N| \geq |b| - 2R$. This implies,
for $|b|R^{-1}$ sufficiently large, that on the support of
$\tilde\Gamma^{R,b}$ (defined in \eqref{deftildegamma}),
\begin{equation}\label{eq:proof1.5}
 V(x_N) + \sum_{i=1}^{N-1} W(x_i - x_N) <
 - \frac{\gamma_0}{|b| + R} + \frac{\gamma_1(N-1)}{|b| - 2R}
 < - \frac{\nu}{2 |b|} ,
\end{equation}
for $\nu = \gamma_0 - \gamma_1(N-1) >0$.
Thus, \eqref{eq:proof1} and \eqref{eq:proof1.5} yield
\begin{equation}\label{eq:proof1.6}
\begin{split}
 \lefteqn{\sum_{\ell=1}^N \langle -\Delta_\ell \tilde\Gamma^{R,b},
 \tilde\Gamma^{R,b}\rangle + \langle I_N(x_1,\ldots, x_n)\tilde\Gamma^{R,b},
 \tilde\Gamma^{R,b}\rangle} & \\
 &  \leq \sum_{\ell = 1}^{N-1} \langle -\Delta_\ell
 \tilde\Psi^{R, N-1}, \tilde\Psi^{R, N-1}\rangle +
 \langle I_{N-1}(x_1, \ldots , x_{N-1}) \tilde\Psi^{R, N-1}
 , \tilde\Psi^{R, N-1}\rangle
 \\
 & \ \ \ +  \langle -\Delta
 \Theta^{R, b}, \Theta^{R, b}\rangle - \frac{\nu}{2|b|} .\\
\end{split}
\end{equation}

Taking into account that $\|\nabla_{x_\ell}\tilde\Psi^{R, N-1}\|
\leq c \| \tilde\Psi^{R, N-1} \|$ ($\ell = 1,\ldots N-1$), and that
$\|\nabla_{x_N} \Theta^{R, b}\| \leq c \|\Theta^{R,b}\|$, with a
constant $c$ independent of $R$, we derive from \eqref{eq:80}
\begin{equation}\label{eq:proof2}
\begin{split}
 \bigg|\sum_{\ell=1}^N \langle\nabla_{x_\ell} A(x_\ell)
 \tilde\Gamma^{R,b}, \tilde\Gamma^{R,b} \rangle
 -  \sum_{\ell=1}^{N-1}
 \langle\nabla_{x_\ell} A(x_\ell) \tilde\Psi^{R,N-1},
 \tilde\Psi^{R,N-1}\rangle \\
 - \langle\nabla_{x_N}
 A(x_N) \Theta^{R,b}, \Theta^{R,b} \rangle \bigg|
 \leq \frac{\varepsilon}{|b| - 2R} .
\end{split}
\end{equation}
Similarly to \eqref{eq:proof2}, and using
\eqref{eq:self-energy3.5}, \eqref{eq:20.5}, \eqref{eq:68}, and
\eqref{eq:68.5}, we have
\begin{equation}\label{eq:proof3}
\begin{split}
 \lefteqn{\sum_{\ell=1}^N \langle A^2(x_\ell)
 \tilde\Gamma^{R,b}, \tilde\Gamma^{R,b} \rangle} & \\
 & \leq  \sum_{\ell=1}^{N-1}
 \langle A^2(x_\ell) \tilde\Psi^{R,N-1},
 \tilde\Psi^{R,N-1}\rangle +
 \langle
 A^2(x_N) \Theta^{R,b}, \Theta^{R,b} \rangle
 + \frac{\varepsilon}{|b| - 2R}
\end{split}
\end{equation}
Along the same lines, we have for the magnetic term, using
\eqref{eq:self-energy4} and \eqref{eq:69},
\begin{equation}\label{eq:proof4}
\begin{split}
 \lefteqn{\sum_{\ell=1}^N \langle \sigma\cdot B(x_\ell)
 \tilde\Gamma^{R,b}, \tilde\Gamma^{R,b} \rangle} & \\
 & \leq \sum_{\ell=1}^{N-1}
 \langle\sigma\cdot B(x_\ell) \tilde\Psi^{R,N-1},
 \tilde\Psi^{R,N-1}\rangle +
 \langle
 \sigma\cdot B(x_N) \Theta^{R,b}, \Theta^{R,b} \rangle
 + \frac{\varepsilon}{|b| - 2R} .
\end{split}
\end{equation}
According to Lemma~\ref{lem:estimate4} we have
\begin{eqnarray}\label{eq:proof5}
 \langle H_f \tilde\Gamma^{R,b}, \tilde\Gamma^{R,b}  \rangle \leq
 \langle H_f \Theta^{R,b}, \Theta^{R,b} \rangle +
 \langle H_f \tilde\Psi^{R, N-1}, \tilde\Psi^{R, N-1}\rangle
 \nonumber\\
 +  c\frac{R^{3/2}}{|b|^{5/2}}
 \langle N_f \tilde\Gamma^{R,b}, \tilde\Gamma^{R,b}\rangle\ .
 \nonumber
\end{eqnarray}
Equality \eqref{eq:normalization} implies that $\langle N_f
\tilde\Gamma^{R,b}, \tilde\Gamma^{R,b}\rangle\ \leq c \left(
\|\tilde\Psi^{R, N-1}\|^2 + \|\Theta^{R,b}\|^2 \right)$

Collecting the estimates \eqref{eq:proof1.6}-\eqref{eq:proof5} we
obtain for any $\varepsilon >0$ and sufficiently large $R$,
\begin{equation}\label{eq:proof6}
 \langle H_{N} \tilde\Gamma^{R,b}, \tilde\Gamma^{R,b}\rangle
 \leq E_{N-1} + \Sigma_0 -\frac{\nu}{2 |b|} +
 \frac{6\varepsilon}{|b| - 2R} + \frac{c R^{3/2}}{|b|^{5/2}}\ .
\end{equation}
To complete the proof of the Theorem, we pick first $R$ large
enough to have $\varepsilon < 48^{-1}\nu$, and then pick $|b|$
sufficiently large to satisfy the inequality $(R |b|^{-1})^{3/2}<
\delta (4c)^{-1}$, which implies
 $$
   \langle H_{N} \tilde\Gamma^{R,b}, \tilde\Gamma^{R,b}\rangle
   < E_{N-1} + \Sigma_0\ .
 $$
\section{Appendix}

\begin{lemma}\label{lem-appendix1}
We define $G_\lambda$ as
 $$
  G_\lambda(y) = \mathcal{F}\left(
  \frac{\varepsilon_\lambda(k)}{|k|^\frac12} \zeta(k) \right)
 $$
where $\mathcal{F}$ denotes the Fourier transform. Then, for
$\lambda=1,2$ and arbitrary $\varepsilon >0$,  $|G_\lambda(y) (1
+|y|) | \in L^{2+\varepsilon}(\R^3)$.
\end{lemma}
\begin{proof}
The statement of the Lemma follows from the Hausdorff-Young
inequality, and the fact that for arbitrarily $\varepsilon>0$,
$\left| \nabla_k \frac{\varepsilon_{\lambda,i}(k)}{|k|^\frac12}
\zeta (k) \right|$ is in $L^{2-\varepsilon}(\R^3)$, for $i=1,2,3$,
which can be checked directly.
\end{proof}

\begin{lemma}\label{lem:loc-estimate-modulusp}
Let $\varphi_1(x)\in H^{1/2}(\R^3)$ with support in the ball of
radius $a R$ centered at the origin, and $\varphi_2(x)\in
H^{1/2}(\R^3)$ with support outside the ball of radius $b R$
centered at the origin. Then for $b > 2a$,
\begin{eqnarray}\label{loc-estimate-modulusp}
 \left|\langle |\nabla| \varphi_1 , \varphi_2
 \rangle \right| \leq
 \frac{1}{3^{1/2}  \pi} \frac{a^{3/2}}{R (b-a)^{5/2}}
 \left(\|\varphi_1\|^2\! +\! \|\varphi_2\|^2\right)
\end{eqnarray}
\end{lemma}
\begin{proof}
Consider the function $u$ defined in \eqref{def-smallu}. Then, for
$\chi_1(x) = u(|x|/(b R))$ and  $\chi_2(x) =
\sqrt{1-\chi_1^2(x)}$, we have, according to
\cite[Theorem9]{LiebYau1988}
\begin{eqnarray*}
\lefteqn{\langle |\nabla| (\varphi_1 + \varphi_2), \varphi_1 +
\varphi_2
 \rangle - \langle |\nabla| \varphi_1, \varphi_1\rangle -
 \langle |\nabla|\varphi_2, \varphi_2\rangle } & & \\
 & \leq & \frac{1}{2\pi^2}\int\int
 \frac{|\varphi_1(x)+ \varphi_2(x)|~|\varphi_1(y) + \varphi_2(y)|}{|x-y|^4} \sum_{i=1,2}
 |\chi_i^2(x) - \chi_i^2(y)|\d y \d y
\end{eqnarray*}
Since $\chi_1 =1$ on the support of $\varphi_1$, $\chi_1=0$ on the
support of $\varphi_2$, we obtain
\begin{eqnarray*}
\lefteqn{\langle |\nabla| (\varphi_1 + \varphi_2), \varphi_1 +
\varphi_2
 \rangle - \langle |\nabla| \varphi_1, \varphi_1\rangle -
 \langle |\nabla|\varphi_2, \varphi_2\rangle } & & \\
 & = & 2\Re \langle |\nabla| \varphi_1, \varphi_2 \rangle \\
 & \leq & \frac{1}{\pi^2}\int\int
 \frac{|\varphi_1(x)|~|\varphi_2(y)|}{|x-y|^4} \d y \d y\\
 & \leq & \frac{2}{\pi 3^{1/2}} \frac{a^{3/2}}{R (b-a)^{5/2}}
 \left(\|\varphi_1\|^2\! +\! \|\varphi_2\|^2\right)
\end{eqnarray*}
\end{proof}

\noindent{\sc Acknowledgements.} 
J.-M. B. and S. V. were financially supported
by the Bayerisch-Franz\"osisches
Hochschulzentrum, and by the European Union through the IHP network
of the EU No. HPRN-CT-2002-00277.
T.C. was supported by a Courant Instructorship.


\bibliographystyle{plain}

\end{document}